\documentclass[a4paper,fleqn]{cas-dc}
\usepackage[numbers]{natbib}

\usepackage{graphicx}
\usepackage{array}
\usepackage{float}
\usepackage{flushend}
\usepackage{subcaption}
\usepackage{appendix}

\def\tsc#1{\csdef{#1}{\textsc{\lowercase{#1}}\xspace}}
\tsc{WGM}
\tsc{QE}
\tsc{EP}
\tsc{PMS}
\tsc{BEC}
\tsc{DE}
\begin{document}
\let\WriteBookmarks\relax
\def\floatpagepagefraction{1}
\def\textpagefraction{.001}

\shorttitle{Selective CO$_2$ reduction}
%\begin{figure*}[h]
%\hspace{-2.70cm}
%\centering
%\includegraphics[scale=0.50]{GA1.pdf}
 %\caption{}
%  \label{pristine1000}
%\end{figure*}
\shortauthors{M. M. Islam and A. Zubair}

\title [mode = title]{Pristine and transition metal doped 2D AlSb as high performance electrocatalyst for selective CO\textsubscript{2} reduction: A first-principles study}                      
%\tnotemark[1,2]

%\tnotetext[1]{This document is the results of the %research
 %  project funded by the National Science %Foundation.}

%\author[1,3]{Mostaqul Islam}[type=editor,
\author[1]{Md. Mostaqul Islam}[orcid=0009-0000-7767-0545]
%\corref{cor1}
                        %[type=editor]
                        %auid=000,bioid=1,
                        %prefix=Md.,
                        %role=Researcher,
                        %orcid=0000-0001-%7511-2910]

%\cormark[1]

%\fnmark[b]

%\ead{mdmostaulislam1160@gmail.com}

%\ead[url]{www.cvr.cc, cvr@sayahna.org}

\credit{Conceptualization, Formal Analysis, Methodology, Visualization, Software, Investigation, Writing – original draft}

\affiliation[1]{organization={ Department of Electrical and Electronic Engineering},
    addressline={Bangladesh University of Engineering and Technology (BUET)}, 
    city={Dhaka},
    % citysep={}, % Uncomment if no comma needed between city and postcode
    postcode={1205}, 
    country={Bangladesh}}

\author[1]{Ahmed Zubair}[orcid=0000-0002-1833-2244]
\corref{cor2}
%\cormark[1]

%\author[1]{Ahmed Zubair\corref{cor1}}
%\author[2]{B. Rahman\corref{cor2}}

%\cortext[cor1]{Corresponding author . 
% E-mail address: 1906070@eee.buet.ac.bd (M. Islam).}
 
\cortext[cor2]{Corresponding author. E-mail address: ahmedzubair@eee.buet.ac.bd (A. Zubair)}

%\author[2,3]{CV Rajagopal}[%
 %  role=Co-ordinator,
% suffix=Jr,
   
%\ead{cvr3@sayahna.org}
%\ead[URL]{www.sayahna.org}

\credit{Supervision, Conceptualization, Methodology, Project administration, Resources, Funding acquisition, Writing – review \& editing}

%\affiliation[2]{organization={Sayahna Foundation},
   % city={Jagathy},
    % citysep={}, % Uncomment if no comma needed between city and postcode
    %postcode={695014}, 
    %state={Trivandrum},
    %country={India}}

%\author%
%[1,3]
%{Rishi T.}
%\cormark[2]
%\fnmark[1,3]
%\ead{rishi@stmdocs.in}
%\ead[URL]{www.stmdocs.in}

%\affiliation[2]{organization={ Department of Electrical and Electronic Engineering},
    %addressline={Bangladesh University of Engineering and Technology}, 
    %city={Dhaka},
    % citysep={}, % Uncomment if no comma needed between city and postcode
   % postcode={695571}, 
    %state={Trivandrum},
    %country={India}}




%edit first three sentences of abstract
\begin{abstract}
Electrochemical CO\textsubscript{2} reduction reaction (CO\textsubscript{2}RR) using %pristine and doped 
2D nanomaterials has emerged as a sophisticated approach to mitigate industrial CO\textsubscript{2} emissions. %while enabling sustainable growth.
%The process of global industrialization has inevitably given rise to CO\textsubscript{2} emission and their adverse impacts. Utilization of this additional CO$_2$ through electrochemical reduction has become an essential technology to be adopted for environmental safety and industrial growth. Both pristine and doped 2D nanomaterials have been reported to be efficient CO$_2$ reduction reaction (CO$_2$RR) electrocatalysts. 
In this work, the potential application of pristine as well as strategically Fe, Co, Ni-doped 2D AlSb was examined as a CO$_2$RR electrocatalyst. %Also, the comparison among top, bottom and hollow site doped 2D AlSb was reported. 
The recation pathways of CO$_2$RR intermediate complexes, %HCOOH, HCHO, CH$_3$OH, CH$_4$,
overpotential, stability, efficiency, and selectivity were studied using %Gibbs free energy calculations and
%first-principles calculations based on 
density functional theory (DFT). Outstanding overpotentials were achieved with pristine and doped 2D AlSb: Ni-doped 2D AlSb was selective for HCOOH (0.12eV) and CH\textsubscript{4} (0.28eV), and Fe-doped 2D AlSb for HCHO (0.31eV) and CH\textsubscript{3}OH (0.31eV). The opposing effects of hydrogen evolution reaction (HER) %[\textit{which is a side reaction}]
was mitigated with the application of electric potential and solution pH. %Electric potential made CO\textsubscript{2} adsorption easier on 2D AlSb, than H\textsubscript{2}. 
The main reasons for the enhancement of catalytic effect due to doping with Fe, Co, and Ni are bandgap reduction and creation of states at the edge of the valence band due to the 3d orbitals of these dopants. Interestingly, the Fe-doped 2D AlSb catalyst exhibited the highest catalytic activity. Excellent electrocatalytic properties of pristine and doped 2D AlSb make them suitable as CO\textsubscript{2}RR catalysts contributing towards a green and sustainable energy ecosystem.
\end{abstract}

% \includegraphics{figs/grabs.pdf}

%\item Explored 2D AlSb as a novel material for CO\textsubscript{2} reduction reaction towards sustainable and green development.
%\item High-performance CO\textsubscript{2} reduction reaction catalyst was designed by anchoring a transition metal on 2D AlSb, which reduced overpotential as well as enhanced efficiency.
%\item Highly stable 2D AlSb with strong selectivity and structural stability confirmed via phonon spectra as well as \textit{ab initio} molecular dynamics simulations.
%\item Enhanced spontaneity of CO\textsubscript{2} reduction reaction by applying electric potential and ensuring favorable reaction medium for 2D AlSb and it's derivatives.

% Keywords

\maketitle

\section{Introduction}
Fossil fuels have been the main energy source for mankind for centuries. However, atmospheric CO$_2$ concentrations have increased significantly with rapid industrialization in recent years. Excessive CO$_2$ emission is a major contributor to the greenhouse effect, leading to global temperature rise, climate change, and ocean acidification \cite{ou2021deep}. Addressing CO$_2$ mitigation is crucial for achieving sustainable and eco-friendly development. Several advanced techniques have been explored to convert CO$_2$ into value-added chemicals, thereby reducing its concentration while utilizing it as a chemical feedstock. Among these approaches, photocatalytic CO$_2$ reduction \cite{wang2021density}, thermocatalytic CO$_2$ reduction \cite{koshy2021bridging}, and electrochemical CO$_2$ reduction \cite{nitopi2019progress} have gained significant attention. CO$_2$ is a chemically inert and stable compound. Hence, the main challenges to select a suitable electrocatalyst are availability, high activity, high selectivity, thermal stability, efficiency, easy fabricability, along with low overpotential, i.e., energy barrier, and low side reaction probability. For the industrial implementation of CO$_2$ reduction, an electrolyzer must achieve a lifespan of approximately 30,000 hours while sustaining a geometric current density above 200mA/cm\textsuperscript{2} \cite{jeanty2018upscaling,fu2010syngas}. However, current researches report significantly lower current densities, often below 30mA/cm\textsuperscript{2}  \cite{ma2013electrochemical}. Additionally, long-term operational stability remains a challenge, with only a few studies demonstrating electrolysis processes exceeding 100 hours. These limitations highlight the need for further advancements to enhance the efficiency and durability of CO$_2$ reduction reaction (CO\textsubscript{2}RR).

Photocatalytic CO$_2$RR provides a comprehensive idea of the reaction mechanisms involved in the photocatalytic reduction of CO\textsubscript{2}. However, a unified mechanistic framework that integrates findings from various studies is still lacking. Most photocatalysts exhibit low quantum efficiency, high recombination rates of photoexcited electron-hole pairs, and low stability under prolonged irradiation. These effects can lead to photodegradation and deactivation \cite{karamian2016general}. %The reaction involves multiple electron transfer steps, complicating intermediate formation and product selectivity in thermocatalytic CO\textsubscript{2}RR. %
Moreover, thermocatalytic CO\textsubscript{2}RR lacks in mechanistic understanding and the design of efficient catalysts for direct conversion of CO\textsubscript{2} into C\textsubscript{1} products thermally. Again, selectivity to products 
%such as %CO\textsubscript{2}, CH\textsubscript{3}OH, CH\textsubscript{4}, and CH\textsubscript{3}-O-CH\textsubscript{3} 
is poor considering both thermodynamic and kinetic perspectives \cite{roy2018thermochemical}. Moreover, many CO\textsubscript{2} hydrogenation reactions are endothermic, necessitating high temperatures, which can lead to unwanted side reactions. Also, traditional metal-based catalysts, e.g., Cu, Ni, Ru, often suffer from sintering and deactivation over time. Moreover, most thermocatalytic CO\textsubscript{2} conversion processes rely on green H\textsubscript{2}, which is currently expensive and energy-intensive to produce. Thermocatalysts undergo coking, sintering, and phase segregation, reducing their efficiency over time.

\begin{figure*}[h]
%\hspace{-2.70cm}
\centering
\includegraphics[scale=0.40]{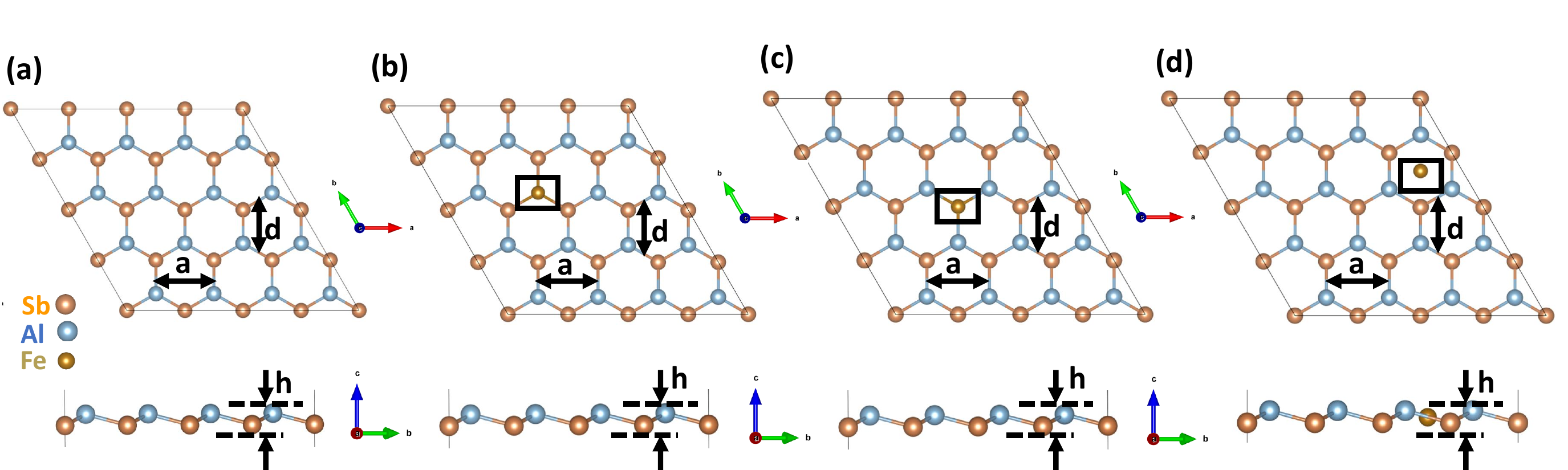}
 \caption{Schematics of the top and side views of (a) pristine, (b) Fe-doped (top site),(c) Fe-doped (bottom site), and (d) Fe-doped (hollow site) 2D AlSb. Here, a is the lattice constant, d is the bond length, and h is the buckling height.}
  \label{pristine1}
\end{figure*}

In the field of electrocatalytic CO$_2$RR, the pioneering work in CO$_2$RR led to significant advances in electrocatalyst development, particularly in exploring various metal electrodes for CO$_2$ conversion \textcolor{black}{\cite{vayenas2008modern}}. Among these, Cu has demonstrated the ability to produce multiple carbon-based products; however, it exhibits notable limitations \cite{popovic2020stability,gattrell2006review}. At the nanoscale, Cu's high sensitivity to reaction conditions, multiple oxidation states, and high overpotential negatively impact its stability, selectivity, and efficiency. Several factors are involved in CO$_2$RR performance, such as the applied electrode voltage, the pH levels of the electrolyte, the surface composition, shape, and structure of the catalyst materials \cite{meng2019rhodium,zhang2021co}. Due to large surface areas, tunable electronic properties, sensitivity to surface flaws, and extraordinary electrical and thermal conductivity \cite{Shahriar2022APSUSC,ghosh2024enhanced}, 2D materials are the focus of most of the CO$_2$RR research. Furthermore, a viable platform for improved CO$_2$RR performance is offered by a 2D supported single-atom catalyst (SAC) \cite{gawande2016cu}. To date, a large number of 2D materials and their derivatives,  including pristine graphene, metal-doped graphene \cite{yang2022theoretical,yang2023theoretical}, pristine phosphorene, Cu-doped phosphorene \cite{zhang2021co}, Ti$_3$C$_2$O$_2$ monolayer \cite{zhou2023efficient}, metal-doped MoS$_2$ \cite{ren2022single}, and Bi$_2$WO$_6$ \cite{liu2022effect} were explored. N\textsubscript{2} and transition metal (TM)-doped graphene have recently been analyzed \cite{yang2022theoretical}. Transitional metal-N\textsubscript{4} (MN\textsubscript{4}), extra graphitic N (MN\textsubscript{4}-GN), pyridinic N 
(MN\textsubscript{4}-PDN), and pyrrolic N (MN\textsubscript{4}-PON)-based graphene as electroreduction catalysts have been prepared physically as 
CO\textsubscript{2}RR electrocatalysts \cite{yang2023theoretical}. However, the overpotential could not be reduced with graphene derivatives. TM single-atom catalyst anchored on 2D Ti\textsubscript{3}C\textsubscript{2}O\textsubscript{2} (TM-Ti\textsubscript{3}C\textsubscript{2}O\textsubscript{2})
\cite{zhou2023efficient} and MoS\textsubscript{2}(M@MoS\textsubscript{2}, M = Fe, Co, Ni, Cu, Ru, Pd and Pt) \cite{ren2022single} as CO\textsubscript{2}RR catalysts were analyzed. A complex material Bi\textsubscript{2}WO\textsubscript{6} (BWO), under solar light irradiation, revealed the efficient CO\textsubscript{2}RR properties \cite{liu2022effect}. However, heterostructures are difficult to synthesize. 2D supported SACs have exhibited several unique features as CO\textsubscript{2}RR catalyst due to tunable coordination environment, quantum confinement, synergy between 2D materials, enhanced stability, and ability to engineer strain. Nonetheless, low overpotential and simple structure SACs are required for efficient CO\textsubscript{2}RR.

III-V semiconducting monolayers have recently emerged as highly promising materials for diverse applications~\cite{Shahriar2022APSUSC}, including in the field of photocatalysis. For instance, halogenated 2D AlSb has demonstrated tremendous photocatalytic performance \cite{shahriar2023halogenation}, while the AlSb /ZnO 2D heterostructure has been established as an efficient water-splitting photocatalyst with significant potential for fuel cell and energy storage applications \cite{ghosh2024enhanced}. However, III-V 2D semiconducting materials have not yet been explored for CO\textsubscript{2}RR. Given their demonstrated broad potential in catalysis, we have selected 2D AlSb as a promising candidate for CO\textsubscript{2}RR applications. To overcome the challenges of electrochemical CO\textsubscript{2}RR, a fundamental understanding of catalytic activity at the atomic level is required, where density functional theory (DFT)-based \cite{yang2022theoretical} calculations are used due to their ability to model complex reaction mechanisms at an atomic scale. These simulations help identify reaction pathways, energy barriers, and active sites. This reduces the need for costly and time-consuming experimental trials while ensuring a targeted approach to catalyst development  {\cite{yang2023theoretical}}.

In this paper, we comprehensively investigated the electrocatalytic effect of novel 2D AlSb and TM (Fe, Co, Ni)-doped 2D AlSb for CO$_2$RR utilizing DFT-based first-principles calculations. First, we calculated the band structure, density of states (DOS), and projected density of states (PDOS) for theoretical insights into AlSb. We performed charge density and Bader charge analysis to further verify the adsorption ability. Next, we studied selectivity by predicting the reaction pathways to CO, HCOOH, CH$_3$OH, HCHO, and CH$_4$ initiated from CO$_2$. We calculated efficiency from the maximum energy barrier (overpotential) required in the most likely pathway. We estimated the energy barriers from Gibbs free energy calculations. We observed thermal and mechanical stability from \textit{ab initio} molecular dynamics (AIMD) calculations. We predicted fabricability and vibrational stability from the phonon-dispersion relation. We used binding energy as the key factor for energetic stability. To determine the photocatalytic capability of the proposed catalyst, we studied optical properties. Finally, we checked the hydrogen evolution reaction (HER) for reaction spontaneity. We conducted a comparative performance analysis of pristine and doped 2D AlSb with previous reports. AlSb-based SACs were proved to be phenomenal for efficient conversion and utilization of CO$_2$ due to tunable bandgap, specific nucleation site, broad surface area, better selectivity and stability. The insights gained from this work will be beneficial for utilizing nanomaterials in eco-friendly CO\textsubscript{2} industrialization, carbon recycling, and the production of renewable fuels.

\begin{figure*}[h]
%\hspace{-2.70cm}
\centering
\includegraphics[scale=0.40]{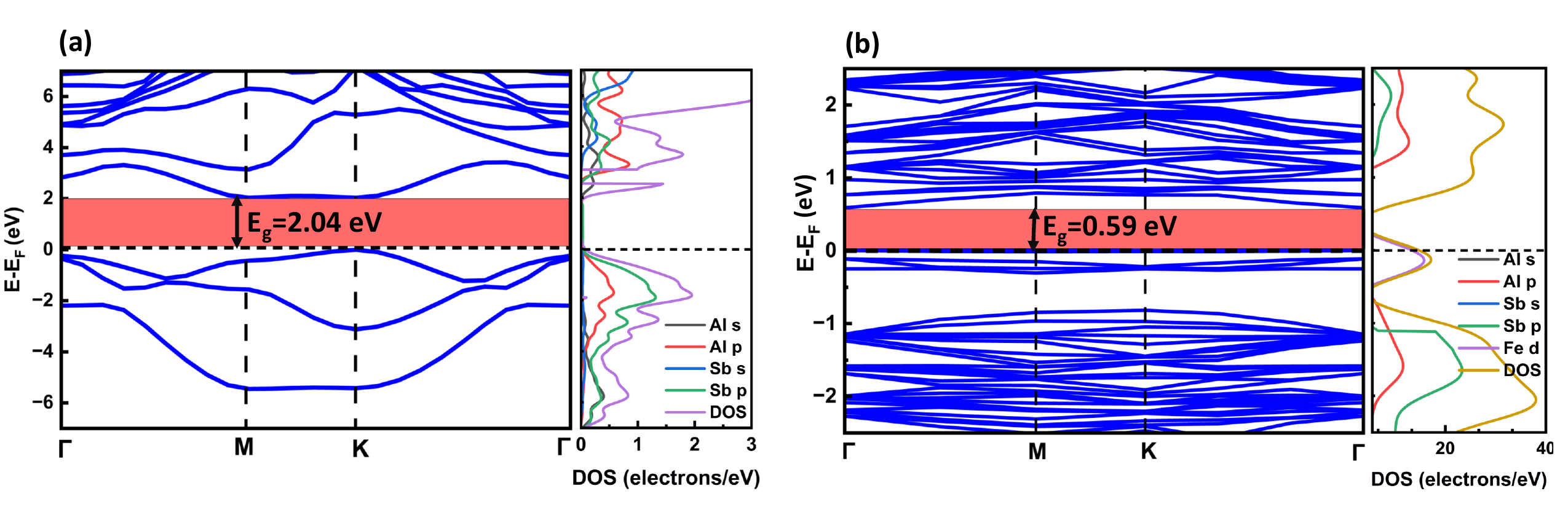}
\caption{Band structure and density of states of (a) unit cell AlSb and (b) Fe-doped (hollow site) 2D AlSb.}
  \label{pristine7}
\end{figure*}

\section{Computational details}

%\subsection {Method of DFT calculation}

All DFT-based calculations were performed using the Cambridge Serial Total Energy Package (CASTEP) \cite{clark2005first} and DMol\textsuperscript{3} modules in Material Studio (MS) software. We constructed a unit cell and a monolayer in CASTEP and optimized them. Afterward, we calculated band structure, DOS, PDOS, and charge density differences (CDD) to extract electronic properties. Additionally, we calculated absorption spectra, permittivity, reflectivity, and transmittance to extract optical properties. We performed simulations for phonon dispersion to determine the stability of the nanostructure. DMol\textsuperscript{3} was used for Gibbs free energy calculations to extract thermodynamic properties. We used Quantum ESPRESSO (QE) for the calculation of the Bader charge on various intermediate complexes. 

The exchange correlation was carried out by Perdew-Burke-Ernzerhof (PBE) along with generalized gradient approximation (GGA) \cite{perdew1996generalized} except for band structure calculation for pristine AlSb, where Heyd-Scuseria-Ernzerhof (HSE06) hybrid functional was used as PBE underestimates bandgap~\cite{borlido2020exchange}. Dispersion interactions were described by the DFT-D3 Grimme method \cite{grimme2010consistent} for relaxation. Ultrasoft pseudopotentials were used for optimization and energy calculations, whereas norm-conserving pseudopotentials were used for phonon dispersion and HSE06 calculations. A cutoff energy of 500 eV was chosen for the plane-wave basis set, and the first Brillouin zone was sampled using a Monkhorst-Pack k-point mesh of (9$\times$9$\times$1). For optimization of the structures, convergence criteria were set to 1.0$\times$10\textsuperscript{-6} eV\AA\textsuperscript{-1} force and 1.0$\times$10\textsuperscript{-6} eV energy on each atom 
where system convergence was set to 1.0$\times$10\textsuperscript{-8} eV/atom for self-consistent field (SCF) calculations. To prevent any interatomic interactions between two periodic layers, a 20{\AA} separation was provided. 

We calculated phonon spectra with the linear response method for vibrational stability and fabricability \cite{shahriar2023halogenation}. And we performed AIMD calculations for mechanical stability %Phonon spectral calculation requires "All bands/EDFT" electronic minimizer and "Fix" occupancy for linear response method.
%We performed AIMD calculations 
on the pristine AlSb using the NPT ensemble at 400 K temperature and 20 atm pressure for 1 ps. We analyzed optical properties using MS for pristine AlSb. 

A (4$\times$4$\times$1) supercell of AlSb was created by repeating the unit cell 4 times in two lateral basis directions of the crystal for doped studies. By calculating the minimum ground state energy, we determined the most active site for doping and created TM (Fe, Co, Ni)-doped 2D AlSb. We optimized the intermediate complexes of CO$_2$RR separately, placed them on AlSb derivatives, and determined the Gibbs free energy of the complexes.

%\subsection {Calculation model}
For energetic stability, the binding energy was calculated.\begin{equation}
    E_b = E_{\text{TM-sub}} - E_{\text{sub}} - E_{\text{TM}}
    \label{eq1}
\end{equation}
Where \( E_{\text{TM-sub}} \) is the energy of TM (Fe, Co, Ni)-doped substrate 2D AlSb, \( E_{\text{sub}} \) is the energy of the substrate without TM,  \( E_{\text{TM}} \) is the energy of isolated TM atom. Negative binding energy implies energetic stability. The more negative the binding energy, the more stable the material is.

The Gibbs free energy change (\(\Delta G\)) for each of the CO\textsubscript{2}RR intermediate complexes was controlled by utilizing a computational hydrogen electrode (CHE) model proposed by N{\o}rskov \textit{et al.} \cite{norskov2004origin} given as

\begin{equation}
    \Delta G = \Delta E + \Delta E_{\text{ZPE}} - T \Delta S + \Delta G_U + \Delta G_{\text{pH}},
\end{equation}

where the \(\Delta E\), \(\Delta E_{\text{ZPE}}\), \(\Delta S\) are the electronic energy differences for each step, the change in zero-point energy and the change in entropy, respectively. \(T\) is the temperature (\(T = 298.15 \text{ K}\)). E, E\textsubscript{ZPE} and \(S\) of CO\textsubscript{2}RR complexes was calculated by applying DFT to the frequencies of vibration. The term \(\Delta G_U\) represents the effect of the applied external potential, which is expressed as \(-neU\), where \(U\) denotes the applied electrode potential and \(n\) is the number of electrons transferred. \(\Delta G_{\text{pH}}\) is the pH correction, which is obtained via the formula, \begin{equation}\Delta G_{\text{pH}} = 2.303 \times k_B T \times \text{pH}
\end{equation}

where \(k_B\) is the Boltzmann constant and the value of pH was set zero in the absence of applied pH.

From the maximum free energy change (\(\Delta G_{\text{max}}\)) between two consecutive reaction intermediate complexes in the most likely pathway, the limiting potential (U$_L$) \cite{zhou2023efficient} of each reaction path to a specific product can be determined. U$_L$ of the potential determining step (PDS) is given as
\begin{equation}
    U_L =\Delta G_{\text{max}} /e.
\end{equation}

\begin{figure*}[h]
%\hspace{-2.70cm}
\centering
\includegraphics[scale=0.40]{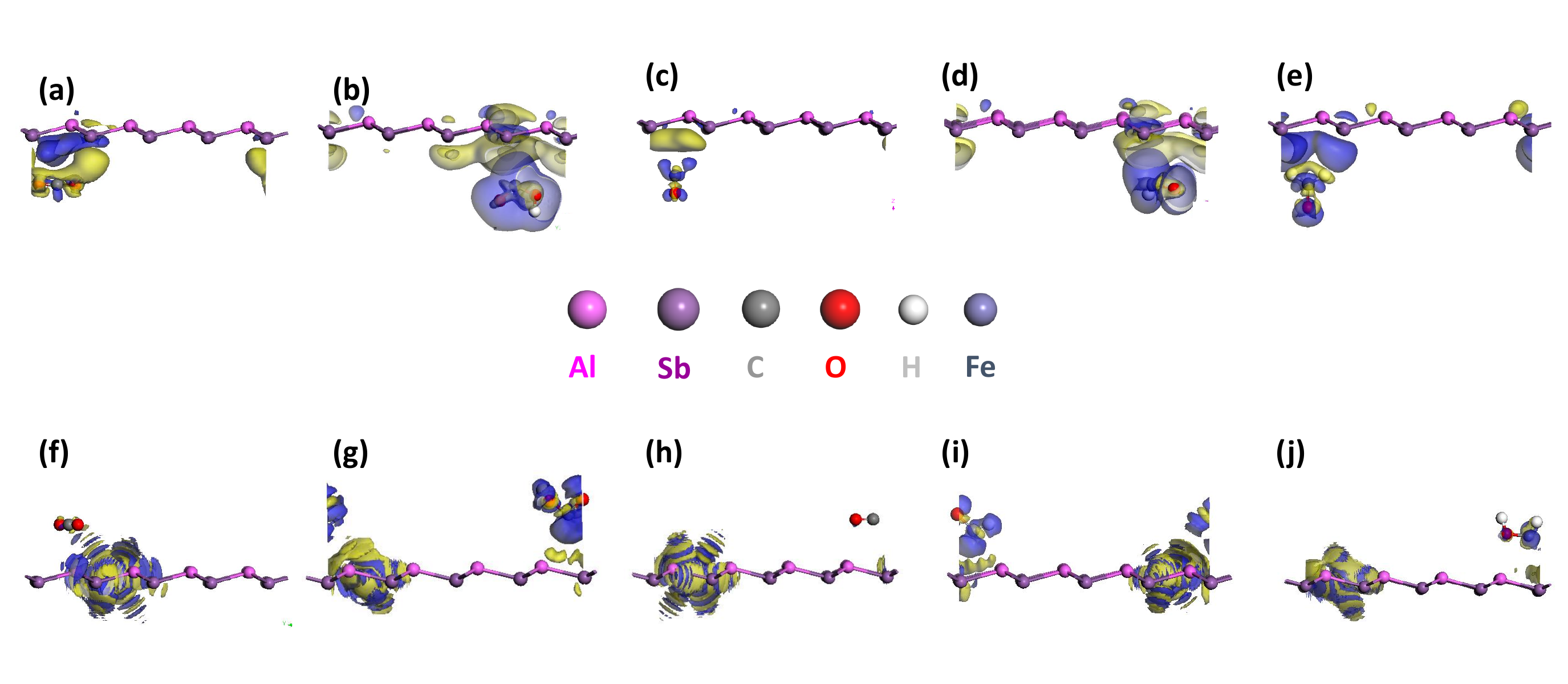}
\caption{Schematics of charge density differences (CDD) for CO\textsubscript{2} to HCHO pathway on pristine 2D AlSb and Fe-doped 2D AlSb. CDD plots of (a)CO\textsubscript{2}, (b)*COOH, (c)CO, (d)*CHO and (e)HCHO on pristine 2D AlSb, respectively. Isosurface=10\textsuperscript{-3} e\textsuperscript{-}\text{\AA}\textsuperscript{-3}. CDD plots of (f)CO\textsubscript{2}, (g)*COOH,  (h)CO, (i)*CHO and (j)HCHO on Fe-doped 2D AlSb, respectively. Isosurface=10\textsuperscript{-2} e\textsuperscript{-}\text{\AA}\textsuperscript{-3}. Blue is e\textsuperscript{-} rich zone and yellow is e\textsuperscript{-} deficit zone.  }
  \label{pristine100}
\end{figure*}

The overpotential, which is the determining parameter of CO\textsubscript{2}RR\,\cite{zhou2023efficient}, can be defined as,
\begin{equation}
    \eta=U_{eq}-U_{L},
    \label{eq5}
\end{equation}
where U$_L$ is the calculated actual limiting potential and U\textsubscript{eq} is the standard equilibrium potentials for half-cell reactions to reduce CO$_2$ into various products at natural temperature and pressure (NTP). For an efficient CO$_2$RR catalyst, the energy barrier of the HER, which is a side reaction, must be greater than the maximum energy barrier of the CO$_2$RR reaction pathway to all the specific products.

\section{Results and discussion}

\subsection{Pristine AlSb unit cell and monolayer}
We created the unit cell of hexagonal AlSb by cleaving the (111) surface of the zinc blende AlSb structure using CASTEP. We calculated geometric, electronic, dynamic, and optical properties using first-principles DFT calculations by the CASTEP tool. The unit cell of hexagonal diatomic AlSb was optimized (both lattice and cell parameters). The lattice constant a, bond length d, and buckling height h were found to be 4.3986 \AA, 2.6190 {\AA}, and 0.6400 {\AA}, 
respectively, indicated in \autoref{pristine1}, which is consistent with the previous report \cite{shahriar2023halogenation}. The calculated structural parameters in \autoref{pristine1} and band structure in \autoref{pristine7} are in
 excellent agreement with the previous theoretical studies \cite{csahin2009monolayer}. We created a (4$\times$4$\times$1) supercell of AlSb, which was utilized as the active surface for CO\textsubscript{2}RR and substrate for the dopants. PDOS shown with band structure in \textcolor{black}{ \autoref{pristine7}} helped to explain the theoretical insights of CO\textsubscript{2}RR with 3d orbital contribution discussed later.

 \begin{table}[h]%[width=.9\linewidth,cols=4,pos=h]
\centering
%\raggedright
%\scriptsize
%\footnotesize
\caption{Binding energies and band gaps for Fe, Co, and Ni-doped AlSb monolayers}\label{tbl1}
\begin{tabular}{cccc}
\toprule
\makecell{\textbf{Dopant}} & 
\makecell{\textbf{Doping Site}} &
\makecell{\textbf{Binding Energy} \\
\textbf{(eV)}} & 
\makecell{\textbf{Bandgap} \\
\textbf{(eV)}} 
%\makecell{\textbf{Binding Energy} \\ \textbf{(eV)}}
\\
\midrule
      & Top & -1.18 & -- \\
Fe & Bottom & -1.11 & -- \\
      & Hollow & -1.26 & 0.59\\
      \hline
      & Top & -0.43 & -- \\
Co & Bottom & -0.45 & -- \\
      & Hollow & -0.53 & 0.76\\
      \hline
      & Top & -0.44 & -- \\
Ni & Bottom & -0.46 & -- \\
      & Hollow & -0.55 & 0.96\\
\bottomrule
\end{tabular}
\end{table}

\subsection{Structural properties of doped AlSb monolayers}
We doped the 2D AlSb with TMs, such as Fe, Co, and Ni, to check catalytic activity improvement. There were basically three types of active sites for any foreign atom to be occupied. (i) Top site (Al atom will be replaced), (ii) bottom site (Sb atom will be replaced), and (iii) hollow site (interatomic space will be occupied). These sites are shown in \autoref{pristine1} and Figs. S1--S2 of the Supplementary Material. We calculated the binding energy of each of the three sites using \autoref{eq1} to determine the energetic stability of the doped structure. The minimum binding energy structure was a hollow site between atoms for all three external embeddings of Fe, Co, and Ni. The results are summarized in \autoref{tbl1}.

%\begin{table}[h]%[width=.9\linewidth,cols=4,pos=h]
%\centering
%\caption{Table of binding energies for Co doped case:}\label{tbl2}
%\begin{tabular}{cc}
%\toprule
%\makecell{\textbf{Doping Site}} & \makecell{\textbf{Binding Energy} \\ \textbf{(eV)}}   %& 
%\makecell{\textbf{Binding Energy} \\ \textbf{(eV)}}
%\\
%\midrule
%Top & -0.43 \\
%Bottom & -0.45 \\
%Hollow & -0.53 \\
%\bottomrule
%\end{tabular}
%\end{table}

%\begin{table}[h]%[width=.9\linewidth,cols=4,pos=h]
%\centering
%\caption{Table of binding energies for Ni doped case:}\label{tbl3}
%\begin{tabular}{cc}
%\toprule
%\makecell{\textbf{Doping Site}} & \makecell{\textbf{Binding Energy} \\ \textbf{(eV)}}   %& 
%\makecell{\textbf{Binding Energy} \\ \textbf{(eV)}}
%\\
%\midrule
%Top & -0.44 \\
%Bottom & -0.46 \\
%Hollow & -0.55 \\
%\bottomrule
%\end{tabular}
%\end{table}

%\begin{table}[h]%[width=.9\linewidth,cols=4,pos=h]
%\centering
%\caption{Table of bandgaps for Fe, Co and Ni doped cases:}\label{tbl4}
%\begin{tabular}{ccc}
%\toprule
%\makecell{\textbf{Doping Sites}} & \makecell{\textbf{Structure}}   & 
%\makecell{\textbf{Band gap} \\ \textbf{(eV)}}
%\\
%\midrule
%Fe & Fe@4x4\_AlSb & 0.59 \\
%Co & Co@4x4\_AlSb & 0.76 \\
%Ni & Ni@4x4\_AlSb & 0.96 \\
%\bottomrule
%\end{tabular}
%\end{table}

\subsection{Electronic properties}
Electronic properties, such as band structure, DOS, and PDOS, of the hexagonal AlSb unit cell and Fe-doped 2D AlSb are presented in \autoref{pristine7}.  The bandgap of AlSb was found to be 2.04 eV, which is consistent with literature \cite{shahriar2023halogenation}. The band gaps of other doped monolayers are listed in {\autoref{tbl1}}. By observing the DOS from {\autoref{pristine7}} and Fig. S4, we observed a peak around the Fermi level in the doped structures, where Fe-doped 2D AlSb had the highest peak, Co-doped 2D AlSb had a medium peak, and Ni-doped 2D AlSb had almost no peak. The PDOS lines explained this phenomenon more comprehensively. From \autoref{pristine7}, the Al-p orbital contributes most of the DOS in the conduction band, and the Sb-p orbital contributes most of the DOS in the valence band in the case of pristine AlSb. However, for the Fe-doped structure, despite Al-p in the conduction band and Sb-p in the valence band, there was 3d orbital contribution from Fe near the Fermi level. This peak arising from the Fe-3d orbital reduced the effective bandgap to 0.59 eV. The same phenomenon occurred in the case of the Co-doped monolayer, where the bandgap reduced to 0.76 eV. The Ni-doped band structure had a minimal 3d contribution near the Fermi level and caused the least reduction in bandgap. The bandgap of Ni-doped 2D AlSb was found to be 0.96 eV. Actually, the band structure figures indicate the 3d orbital electronic effect maximum for Fe, moderate for Co, and minimum for Ni. Hence, the reason for a huge bandgap reduction for the Fe-doped structure was a strong 3d orbital interaction with Al-p in the conduction band and Sb-p in the valence band from the hollow site. This phenomenon was verified in the projected band structures of Fe-doped 2D AlSb presented in Figs. S5-S8. Amount of decrement in bandgap was reflected in increased CO\textsubscript{2}RR catalytic effect, which will be explained in the later sections. So, 3d orbital interaction and bandgap reduction were the two main factors for CO\textsubscript{2}RR catalytic effect. Band structures of Co and Ni-doped 2D AlSb are shown in Fig. S4.
%\clearpage

\subsection{Charge density analysis for CO\textsubscript{2}RR catalyst}
CDD calculations were performed on the CO\textsubscript{2}RR intermediate complexes adsorbed on pristine 2D AlSb and Fe-doped 2D AlSb. The CDD plot demonstrates charge transfer from the substrate to the dopants analytically. This charge transfer facilitated the effect of CO\textsubscript{2}RR catalysis. The mathematical formula for CDD is given by,
\begin{equation}
  \Delta \rho = \rho_{Substrate+Adsorbate} - \rho_{Substrate} -\rho_{Adsorbate}.
  \label{eq9}
\end{equation}
From \autoref{pristine100}(a), we can observe that, e\textsuperscript{-} cloud (blue region) is being shifted from 2D AlSb to CO\textsubscript{2}. \autoref{pristine100}(b) shows more electronic charges accumulated by the adsorbate *COOH and more electronic charges depleted (yellow region) from the substrate. \autoref{pristine100}(c) shows the CO charge density is completely isolated from the substrate, which indicates CO is a stable product. Again, \autoref{pristine100}(d) shows stronger electronic charges accumulated by adsorbate *CHO and electronic charges depleted from the substrate. \autoref{pristine100}(e) demonstrates another stable product, HCHO, with a charge density region separated from 2D AlSb. The CDD plots validated 2D AlSb to be a CO\textsubscript{2}RR catalyst more intuitively. \autoref{pristine100}(f) to \autoref{pristine100}(j) exhibit CDD on Fe-doped 2D AlSb with the same intermediate complexes. Charges were accumulated in two regions: (i) where Fe was doped and (ii) where intermediate complexes were adsorbed. Clearly, the charge between Fe and Al indicated a bond (bond length 2.535\AA), as well as the charge between Fe and Sb indicated another bond (bond length 2.528 \AA). Bader charge calculations were also performed to examine the amount of transferred charge quantitatively and are shown in \autoref{tbl11}.

\begin{table}[h]%[width=.9\linewidth,cols=4,pos=h]
\centering
\caption{Transferred charge on different intermediate complexes of CO\textsubscript{2}RR}\label{tbl11}
\resizebox{0.95\linewidth}{!}{
\begin{tabular}{cccc}
\toprule
\makecell{\textbf{Intermediate}} \\ \textbf{Complexes} & \makecell{\textbf{Charge on C}}  & 
\makecell{\textbf{Charge on O}} &
\makecell{\textbf{Charge on H}} \\
\midrule
CO\textsubscript{2} & +0.000095e\textsuperscript{-} & -0.023468e\textsuperscript{-} & -- \\
*COOH & -0.164841e\textsuperscript{-} & -0.088019e\textsuperscript{-} & -0.027431e\textsuperscript{-} \\
CO & -0.024155e\textsuperscript{-} & -0.006269e\textsuperscript{-} & --\\
*CHO & -0.066809e\textsuperscript{-} & -0.055631e\textsuperscript{-} & -0.025127e\textsuperscript{-}\\
HCHO & -0.027562e\textsuperscript{-} & -0.009630e\textsuperscript{-} & -0.004715e\textsuperscript{-}\\
\bottomrule
\end{tabular}}
\end{table}

\subsection{Stability}
\subsubsection{Energetic stability}
We determined energetic stability from highly negative binding or adsorption energies. The binding energy was calculated by using \autoref{eq1}. If E\textsubscript{b} from {\autoref{eq1}} is negative, then the structure is assumed to be energetically stable. From {\autoref{tbl1}}, hollow site doped substrates have the lowest binding energies. Therefore, the hollow site was the most energetically favorable structure. That's why CO\textsubscript{2} and other intermediate complexes were easily adsorbed and reduced at hollow site doped structures.

\begin{figure*}[H]
%\hspace{-2.70cm}
\centering
\includegraphics[scale=0.52]{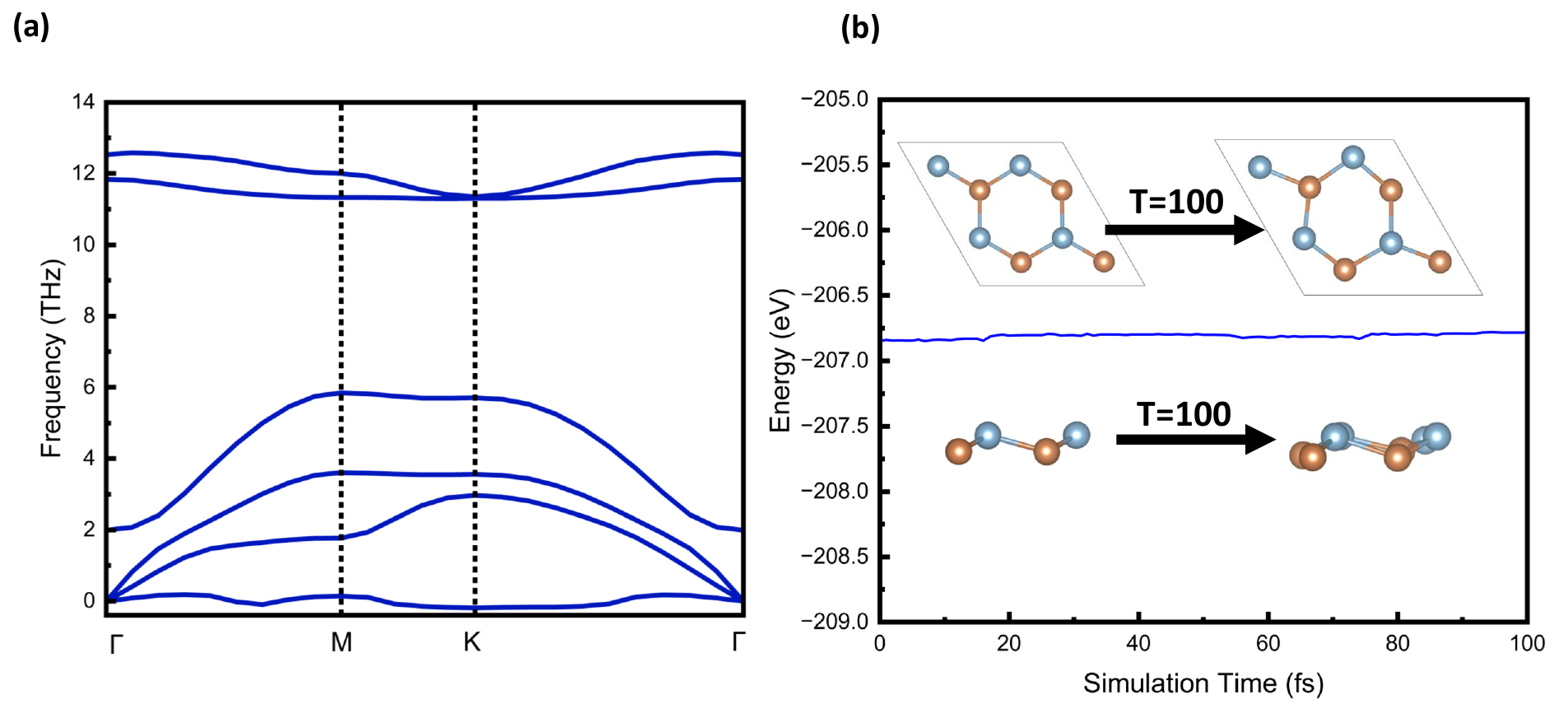}
%\caption{Hydrogen evolution reaction (HER) on (a) pristine AlSb, (c) Fe doped, (e) Co doped (g) ,Ni doped AlSb and modified pathway under applied potential on (b) pristine AlSb, (d) Fe doped, (f) Co doped, (h) Ni doped AlSb respectively.}
  \label{pristine25}
\end{figure*}

\begin{figure*}[H]
%\hspace{-2.70cm}
\centering
\includegraphics[scale=0.52]{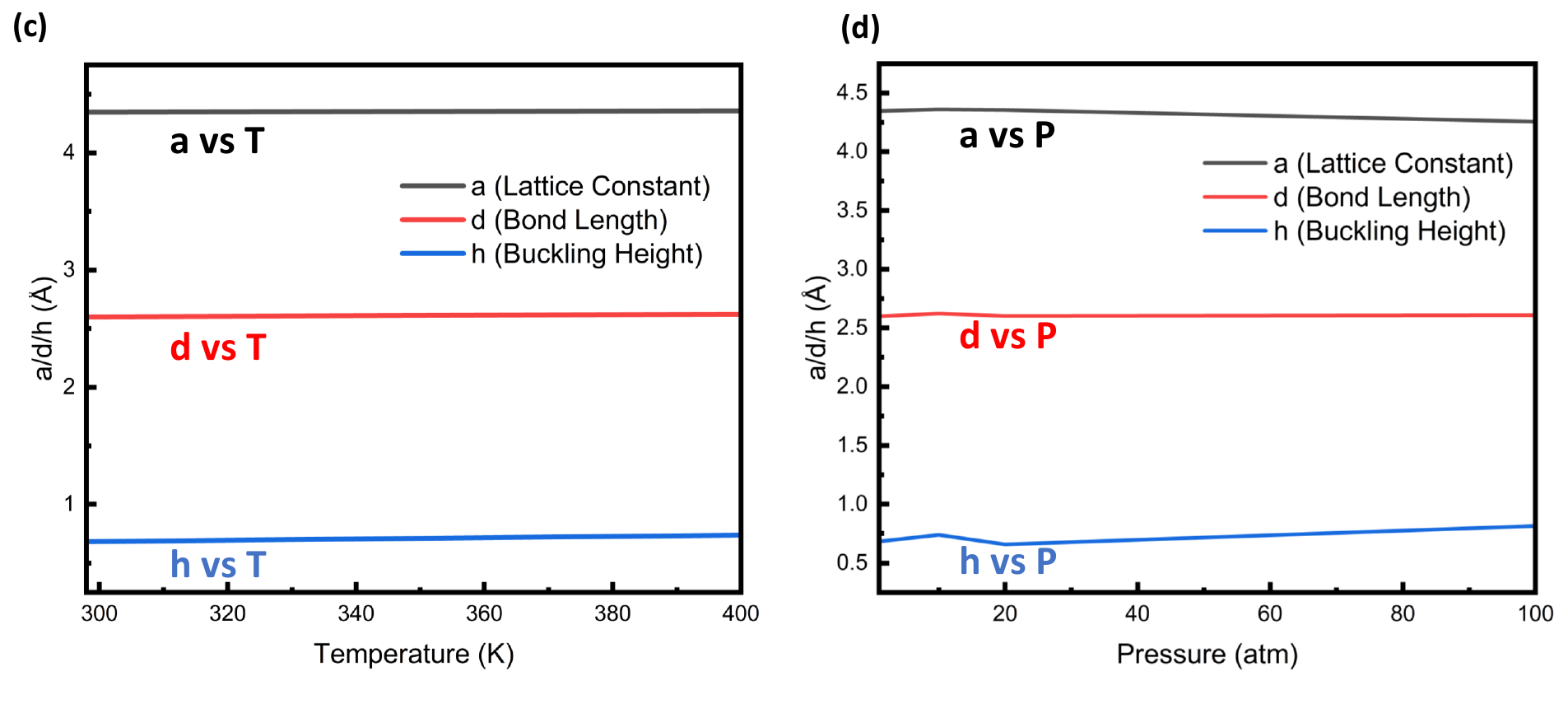}
\caption{(a) Phonon dispersion curves of AlSb, (b) AIMD simulation with NPT ensemble and T = 400 K, P = 20 atm, timestamp = 1 fs. Variation of the lattice constant, a, bond length, d, and buckling height, h, with respect to (c) temperature and (d) pressure.}
  \label{pristine26}
\end{figure*}

\subsubsection{Vibrational stability and fabricability}
We calculated the phonon dispersion relation or the phonon spectra of the pristine hexagonal AlSb monolayer, and the output is exhibited in {\autoref{pristine26}}(a). The linear response method, also known as density functional perturbation theory (DFPT), was employed for the calculation of phonon spectra, as it was computationally simple and cheaper. Two types of vibration modes are clearly observed, (i) acoustic mode and (ii) optical mode from {\autoref{pristine26}}(a). There is a negligible amount of non-positive frequencies in the acoustic mode. In other words, there is hardly any imaginary frequency of vibration that can decrease the energy of the structure and transform it into some other material. The hexagonal AlSb \cite{herczog1958preparation} can be assumed to be vibrational-stable and physically synthesizable \textcolor{black}{\cite{shahriar2023halogenation}}. 

\subsubsection{Mechanical stability}
For mechanical stability \cite{korznikova2017instability}, we performed AIMD simulations, and results are shown in {\autoref{pristine26}}(b). The dynamic energy of the AlSb monolayer during AIMD simulations was almost a constant with the NPT ensemble up to T = 400 K and P = 0.02 GPa (20 atm). The calculations were performed for 100 fs with a timestamp of 1 fs. Though the time duration may seem very small, this duration was for both the high-temperature ramp and the high-pressure ramp simultaneously. Both pressure and temperature, with a higher time duration, are extremely costly and increase complexity. This type of ramp is essential in fabrication processes, and constant energy implies high stability during these impulsing temperature and pressure, which is critical.

Also, after 100 fs of high-temperature and -pressure simulations, there were almost no distortions in 2D AlSb. As shown in {\autoref{pristine26}}(c) and {\autoref{pristine26}}(d), characteristic lattice lengths (a, d, and h) are almost unchanged with respect to high temperature up to 400 K and high pressure up to 100 atm. These two graphs firmly confirmed the high mechanical stability of 2D AlSb under elevated temperature and pressure.

\subsection{Mechanism and selectivity to different products of CO\textsubscript{2}RR }
The six factors required for understanding CO\textsubscript{2}RR mechanism are: (i) source gas CO\textsubscript{2}, (ii) reduction gas H\textsubscript{2}, (iii) catalytic material or catalyst in the form of a solid device, (iv) potential control system, (v) pH control system, and (vi) doping system. CO\textsubscript{2} was adsorbed by the catalyst due to p orbital interaction from Al, Sb, and 3d orbital interaction from Fe, Co, and Ni. Then, this carbonated catalyst became highly active and adsorbed H\textsubscript{2} (H\textsuperscript{+}+e\textsuperscript{-}). After multiple reduction steps, there were some stable products that are environmentally friendly, such as HCOOH, HCHO, CH\textsubscript{3}OH, CH\textsubscript{4}. Selectivity to a specific product was mainly controlled by voltage, pH, and dopants.

\subsubsection{Reaction pathway analysis}
The complete CO\textsubscript{2}RR was broken down into eight electron steps. \textit{Here * denotes the product that is not stable. HWO means hydrogenation with O\textsubscript{2} and HWC means hydrogenation with C}. The intermediate complexes adsorbed on Fe-doped 2D AlSb are shown in Fig. S3. The intermediate complexes of different steps were taken mainly from the previous literature \cite{zhou2023efficient,ren2022single}. The reactions are given below:

\begin{flushleft}
\textbf{Step 1:}
\end{flushleft}
\begin{flushleft}
\text{CO}\textsubscript{2} + \text{H}\textsuperscript{+} + e\textsuperscript{-}= \text{*COOH} \text{ (HWO)} \\
\text{CO}\textsubscript{2} + \text{H}\textsuperscript{+} + e\textsuperscript{-}= \text{*OCHO} \text{ (HWC)}
\end{flushleft}

\begin{flushleft}
\textbf{Step 2:}
\end{flushleft}
\begin{flushleft}
  \text{*OCHO} + \text{H}\textsuperscript{+} + e\textsuperscript{-}= \text{ HCOOH} \text{ (HWO)} \\
  \text{*OCHO} + \text{H}\textsuperscript{+} + e\textsuperscript{-}= \text{*CH}\textsubscript{2}\text{O}\textsubscript{2} \text{        (HWC)} \\
  \text{*COOH} + \text{H}\textsuperscript{+} + e\textsuperscript{-}=\text{CO} + \text{H}\textsubscript{2}\text{O} \text{ (HWO)}\\
\text{*COOH} + \text{H}\textsuperscript{+} + e\textsuperscript{-}= \text{HCOOH} \text{ (HWC)}
\end{flushleft}

\begin{flushleft}
\textbf{Step 3:}
\end{flushleft}
\begin{flushleft}
    \text{CO} + \text{H}\textsuperscript{+} + e\textsuperscript{-}= \text{*COH} \text{ (HWO)}\\  
\text{CO} + \text{H}\textsuperscript{+} + e\textsuperscript{-}= \text{*CHO} \text{ (HWC)}\\
\text{HCOOH} + \text{H}\textsuperscript{+} + e\textsuperscript{-}= \text{*CHOHOH} \text{ (HWO)} \\ 
\text{HCOOH} + \text{H}\textsuperscript{+} + e\textsuperscript{-}= \text{*OCH}\textsubscript{2}\text{OH} \text{ (HWC)} \\
\text{*CH}\textsubscript{2}\text{O}\textsubscript{2} + \text{H}\textsuperscript{+} + e\textsuperscript{-}=\text{*OCH}\textsubscript{2}\text{OH} \text{ (HWO)}
\end{flushleft}

\begin{flushleft}
\textbf{Step 4:}
\end{flushleft}
\begin{flushleft}
\text{*COH} + \text{H}\textsuperscript{+} + e\textsuperscript{-}= \text{*C} + \text{H}\textsubscript{2}\text{O} \text{ (HWO)}\\
\text{*COH} + \text{H}\textsuperscript{+} + e\textsuperscript{-}=\text{*CHOH} \text{ (HWC)}\\
\text{*CHO} + \text{H}\textsuperscript{+} + e\textsuperscript{-}= \text{*CHOH} \text{ (HWO)} \\
\text{*CHO} + \text{H}\textsuperscript{+} + e\textsuperscript{-}=\text{HCHO} \text{ (HWC)}\\
\text{*CHOHOH} + \text{H}\textsuperscript{+} + e\textsuperscript{-}= \text{*CHOH} + \text{H}\textsubscript{2}\text{O} \text{ (HWO)}\\
\text{*OCH}\textsubscript{2}\text{OH} + \text{H}\textsuperscript{+} + e\textsuperscript{-}=\text{HCHO} + \text{H}\textsubscript{2}\text{O} \text{ (HWO)}
\end{flushleft}

\begin{flushleft}
\textbf{Step 5:}
\end{flushleft}
\begin{flushleft}
    \text{*C} + \text{H}\textsuperscript{+} + e\textsuperscript{-}=\text{*CH} \text{ (HWC)}\\
\text{*CHOH} + \text{H}\textsuperscript{+} + e\textsuperscript{-}=\text{*CH} + \text{H}\textsubscript{2}\text{O} \text{ (HWO)} \\ 
\text{*CHOH} + \text{H}\textsuperscript{+} + e\textsuperscript{-}= \text{*CH}\textsubscript{2}\text{OH} \text{ (HWC)}
\end{flushleft}

\begin{flushleft}
\textbf{Step 6:}
\end{flushleft}
\begin{flushleft}
\text{*CH} + \text{H}\textsuperscript{+} + e\textsuperscript{-}=\text{*CH\textsubscript{2}} \text{ (HWC)}\\
\text{*CH}\textsubscript{2}\text{OH} + \text{H}\textsuperscript{+} + e\textsuperscript{-}= \text{*CH}\textsubscript{2} + \text{H}\textsubscript{2}\text{O} \text{ (HWO)} \\  
\text{*CH}\textsubscript{2}\text{OH} + \text{H}\textsuperscript{+} + e\textsuperscript{-}= \text{CH}\textsubscript{3}\text{OH} \text{ (HWC)}  
\end{flushleft}

\begin{flushleft}
\textbf{Step 7:}
\end{flushleft}
\begin{flushleft}
   \text{*OCH}\textsubscript{3} + \text{H}\textsuperscript{+} + e\textsuperscript{-}=\text{CH}\textsubscript{3}\text{OH} \text{ (HWO)}\\  
\text{*OCH}\textsubscript{3} + \text{H}\textsuperscript{+} + e\textsuperscript{-}= \text{CH}\textsubscript{4} + \text{*O} \text{ (HWC)}\\
\text{*CH\textsubscript{2}} + \text{H}\textsuperscript{+} + e\textsuperscript{-}=\text{*CH\textsubscript{3}} \text{ (HWC)}
\end{flushleft}

\begin{flushleft}
\textbf{Step 8:}
\end{flushleft}
\begin{flushleft}
    \text{*CH\textsubscript{3}} + \text{H}\textsuperscript{+} + e\textsuperscript{-}=\text{CH\textsubscript{4}} \text{ (HWC)}
\end{flushleft}
In the first step, CO\textsubscript{2} was reduced to form *COOH or *OCHO. *COOH was formed if O atom was reduced by H\textsubscript{2}, and *OCHO was formed if C atom was reduced by H\textsubscript{2}. We defined these two terms as HWO and HWC. In the second step, *COOH or *OCHO was reduced. For *OCHO, HWO and HWC produced HCOOH and *CH\textsubscript{2}O\textsubscript{2} respectively. For *COOH, HWO and HWC produced CO and HCOOH, respectively. In the third step, CO or HCOOH or *CH\textsubscript{2}O\textsubscript{2} was reduced. For CO, HWO and HWC produced *COH and *CHO, respectively. For HCOOH, HWO and HWC produced *CHOHOH and *OCH\textsubscript{2}OH, respectively. For *CH\textsubscript{2}O\textsubscript{2}, HWO produced *OCH\textsubscript{2}OH only. In the fourth step, *COH or *CHO or *CHOHOH* or *OCH\textsubscript{2}OH was reduced. For *COH, HWO and HWC produced *C and *CHOH, respectively. For *CHO, HWO and HWC produced *CHOH and HCHO, respectively. For *CHOHOH, HWO produced *CHOH only. For *OCH\textsubscript{2}OH, HWO produced HCHO only.  In the fifth step, *C or *CHOH was reduced. For *C, HWC produced *CH only. For *CHOH, HWO and HWC produced *CH and *CH\textsubscript{2}OH, respectively. In the sixth step, *CH or *CH\textsubscript{2}OH was reduced. For *CH, HWC produced *CH\textsubscript{2} only. For *CH\textsubscript{2}OH, HWO and HWC produced *CH\textsubscript{2} and CH\textsubscript{3}OH, respectively. In the seventh step, *OCH\textsubscript{3} or *CH\textsubscript{2} was reduced. For *OCH\textsubscript{3}, HWO and HWC produced *CH\textsubscript{3}OH and CH\textsubscript{4}, respectively. For *CH\textsubscript{2}, HWC produced *CH\textsubscript{3} only. In the eighth and final step, *CH\textsubscript{3} was reduced and HWC produced CH\textsubscript{4} only. 

%\clearpage
\begin{figure*}[h]
%\hspace{-2.70cm}
\centering
\includegraphics[scale=0.49]{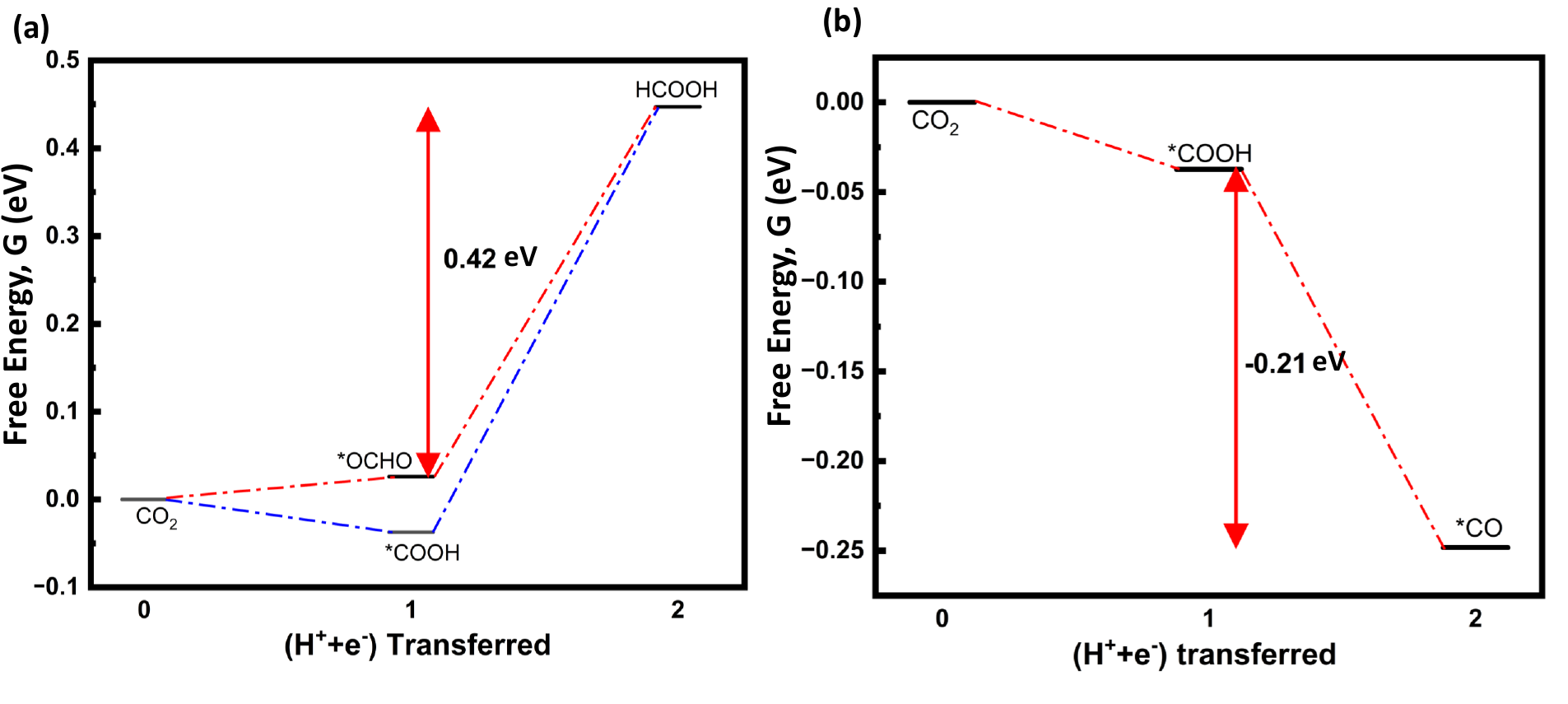}
% \caption{Band structure and density of states of Ni doped(hollow site) AlSb monolayer.}
  \label{pristine9}
\end{figure*}

%\vspace{-0.5cm}
%\hfill
\begin{figure*}[h]
%\hspace{-2.70cm}
\centering
\includegraphics[scale=0.5]{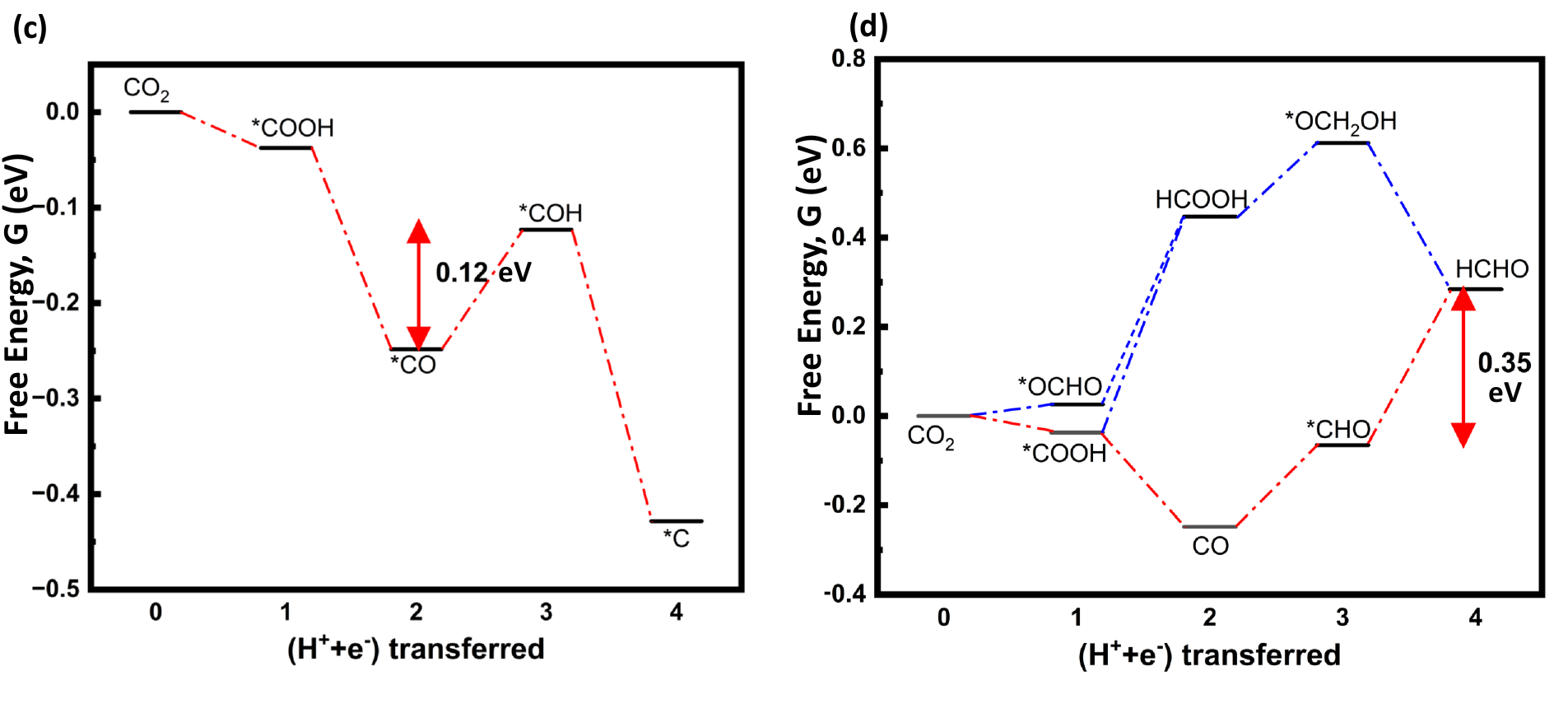}
% \caption{Band structure and density of states of Ni doped(hollow site) AlSb monolayer.}
  \label{pristine10}
\end{figure*}
%hfill
\begin{figure*}[h]
%\hspace{-2.70cm}
\centering
\includegraphics[scale=0.5]{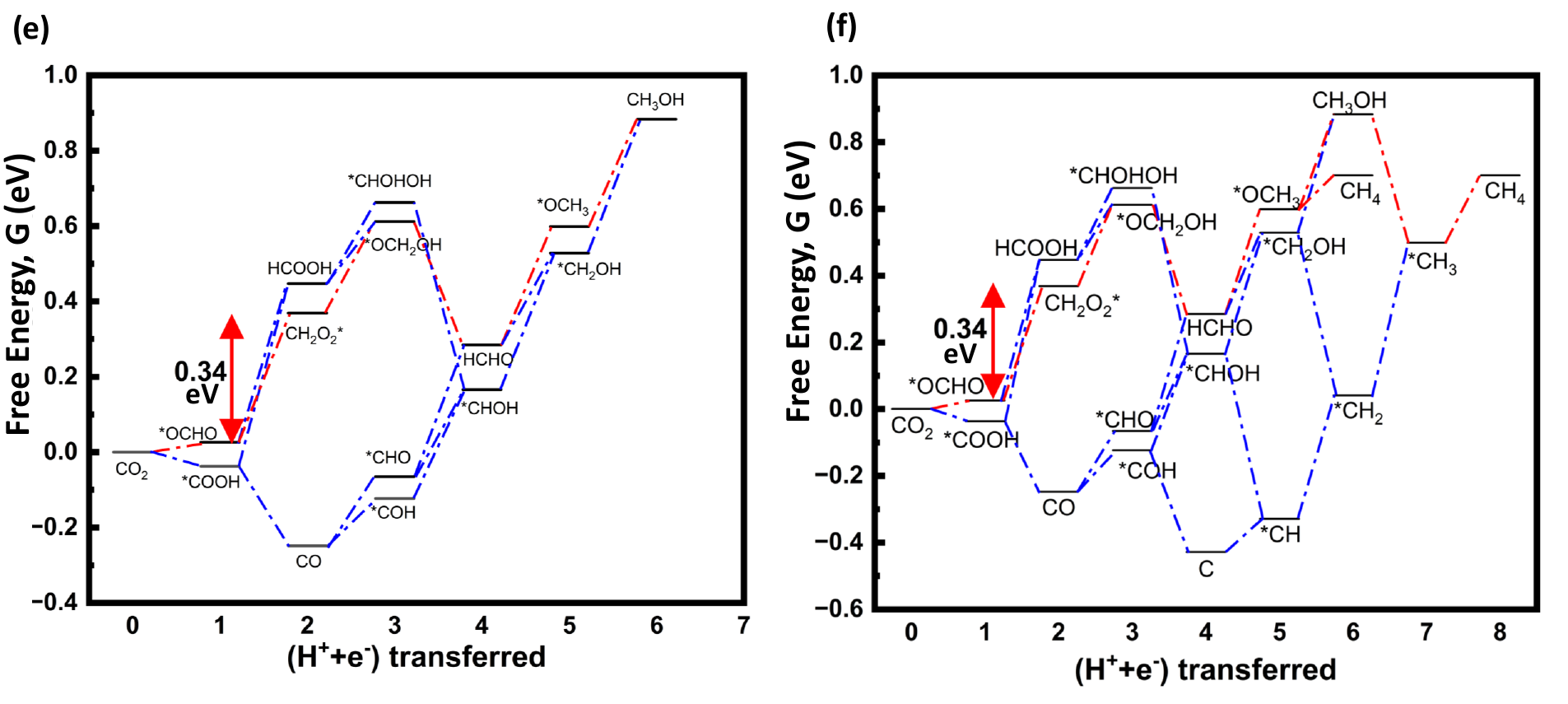}
\caption{Reaction pathway for CO\textsubscript{2} reduction on 2D AlSb to (a) HCOOH, (b) CO, (c) *C, (d) HCHO, (e) CH\textsubscript{3}OH, (f) CH\textsubscript{4}.}
  \label{pristine11}
\end{figure*}
%\clearpage
\begin{figure*}[h]
%\hspace{-2.70cm}
\centering
\includegraphics[scale=0.484]{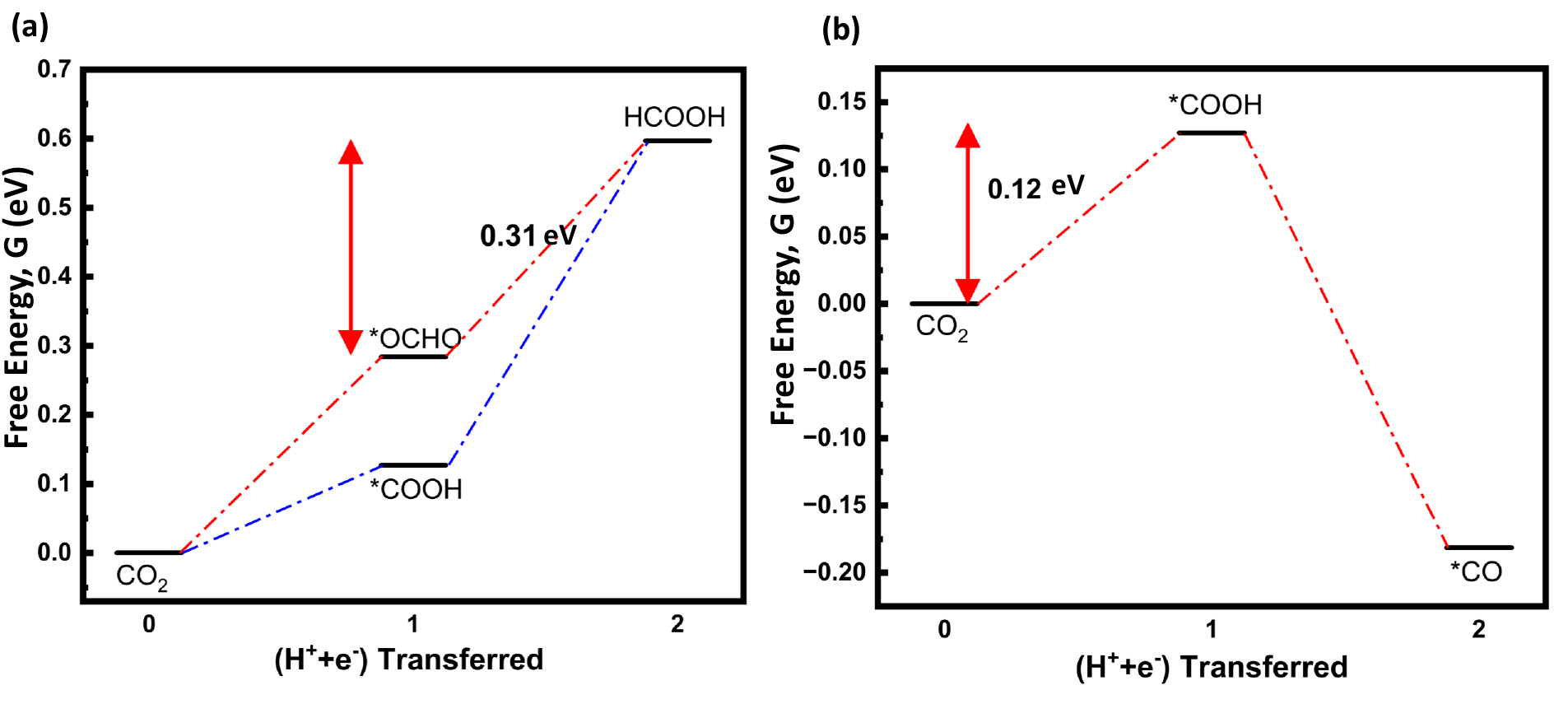}
% \caption{Band structure and density of states of Ni doped(hollow site) AlSb monolayer.}
  \label{pristine12}
\end{figure*}
\begin{figure*}[h]
%\hspace{-2.70cm}
\centering
\includegraphics[scale=0.5]{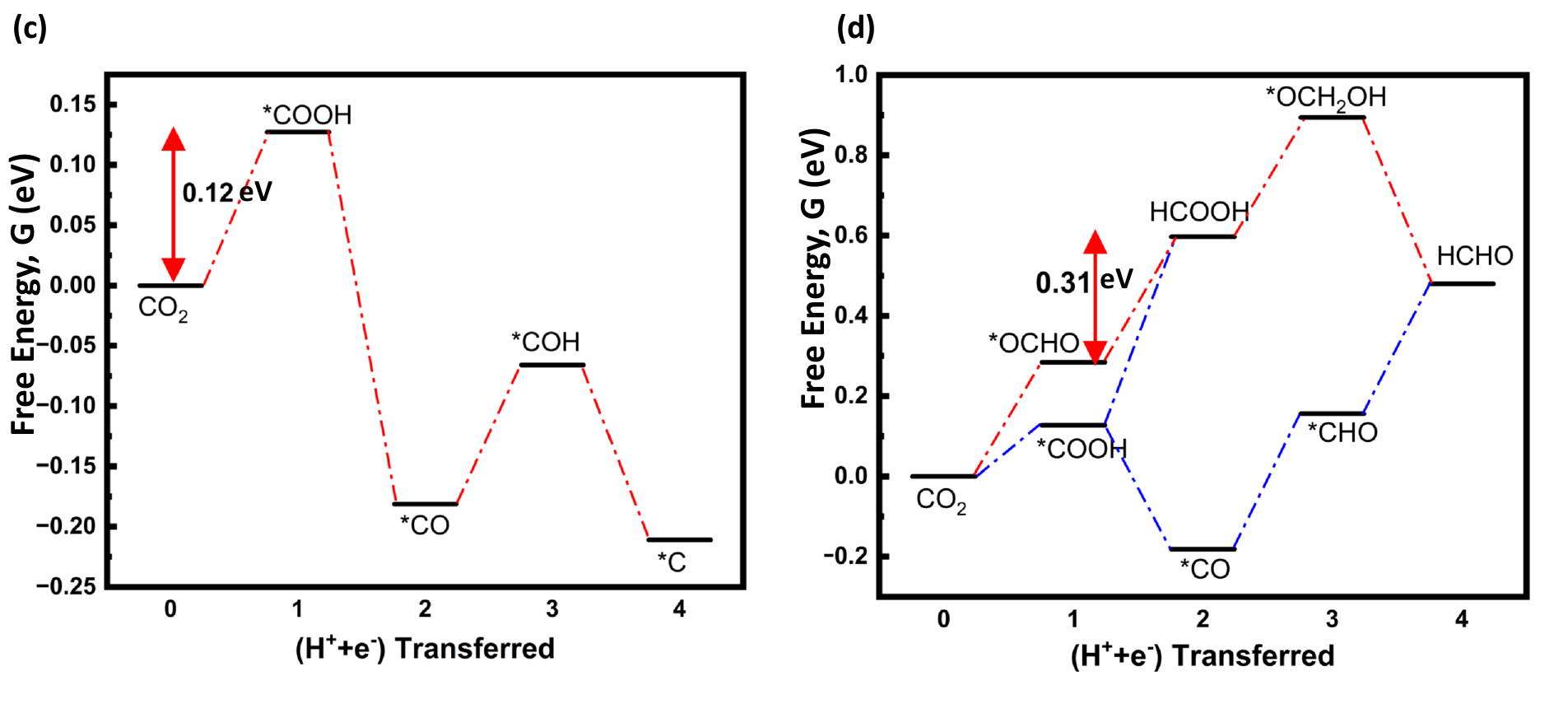}
% \caption{Band structure and density of states of Ni doped(hollow site) AlSb monolayer.}
  \label{pristine13}
\end{figure*}
\begin{figure*}[h]
%\hspace{-2.70cm}
\centering
\includegraphics[scale=0.5]{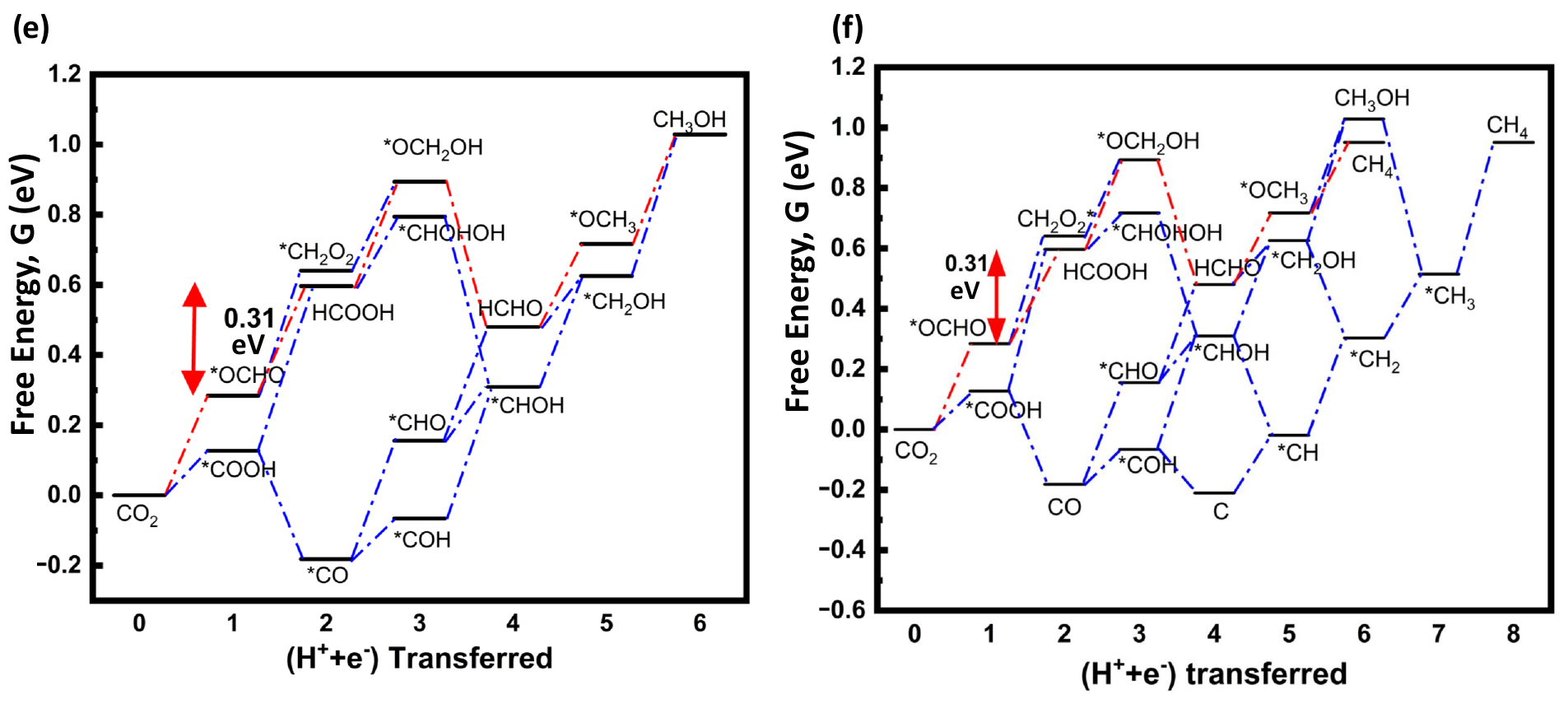}
\caption{Reaction pathway for CO\textsubscript{2} reduction on Fe-doped 2D AlSb to (a) HCOOH, (b) CO, (c) *C, (d) HCHO, (e) CH\textsubscript{3}OH, (f) CH\textsubscript{4}.}
  \label{pristine14}
\end{figure*}

\subsubsection{Selectivity to HCOOH}
First, the term U\textsubscript{L} from \autoref{eq5} needs to be defined. U\textsubscript{L} is the maximum energy barrier of the most likely pathway. The most likely pathway is the pathway that has the lowest limiting potential. U\textsubscript{eq} was considered to be zero, and the value of U\textsubscript{L} was taken as the value of overpotential ($\eta$) according to \autoref{eq5}. Therefore, $\eta$ was used interchangeably with U\textsubscript{L} in this paper. \textit{The red-marked path is the most likely pathway in all diagrams, and the red bidirectional arrow denotes overpotential.} In {\autoref{pristine11}}, the most likely pathway from CO\textsubscript{2} to HCOOH is from CO\textsubscript{2} to *OCHO and then *OCHO to HCOOH which is indicated by red line. Because in the first step, though *COOH formation was more likely (less energy is required), the pathway from *COOH to HCOOH was not chosen as this path required a higher overpotential of about 0.50 eV in the second step, where the pathway from *OCHO to HCOOH required a 0.42 eV maximum barrier. Hence, *OCHO to HCOOH was the potential-determining step (PDS) here, and overpotential from CO\textsubscript{2} to HCOOH conversion was 0.42 eV for pristine 2D AlSb. Similarly, for the pathway to HCOOH in {  \autoref{pristine14}}(a) shown for Fe-doped AlSb. Here, *OCHO to HCOOH was also the PDS, and overpotential from CO\textsubscript{2} to HCOOH conversion for Fe-doped 2D AlSb was 0.31 eV, which was a significant improvement over that without doping. Moreover, the pathway to HCOOH for Co-doped AlSb is shown in Fig. S9(a). Similarly, *OCHO to HCOOH was the PDS and overpotential from CO\textsubscript{2} to HCOOH conversion for Co-doped 2D AlSb was 0.40 eV, which was a slight improvement than that of without doping. Reaction pathways of Co@AlSb and Ni@AlSb are shown in Figs. S9-S14. Again, the pathway to  HCOOH for Ni-doped AlSb is shown in Fig. S12(a). Here, CO\textsubscript{2} to *COOH was the PDS and overpotential from CO\textsubscript{2} to HCOOH conversion for Ni-doped 2D AlSb was 0.12 eV, which was an extraordinary improvement over that of without doping. All four cases -- pristine, Fe-doped, Co-doped, and Ni-doped AlSb -- are tabularized in {\autoref{tbl5}}. Hence, a low overpotential catalyst could be chosen by varying the dopant material.

%\clearpage
\begin{table}[h]%[width=.9\linewidth,cols=4,pos=h]
\centering
\scriptsize
\caption{PDS and overpotential for different catalysts to 2e\textsuperscript{-} product HCOOH, 4e\textsuperscript{-} product HCHO, 6e\textsuperscript{-} product CH\textsubscript{3}OH, 8e\textsuperscript{-} product CH\textsubscript{4}}\label{tbl5}
\begin{tabular}{cccc}
\toprule
\makecell{\textbf{Product}} &
\makecell{\textbf{Catalyst}} & \makecell{\textbf{PDS}}  & 
\makecell{\textbf{Overpotential, $\eta$} \\ \textbf{(eV)}}
\\
\midrule
 & Pristine AlSb & *OCHO $\rightarrow$ HCOOH & 0.42 \\
HCOOH &Fe@AlSb & *OCHO $\rightarrow$ HCOOH & 0.31 \\
 &Co@AlSb & *OCHO $\rightarrow$ HCOOH & 0.40 \\
 &Ni@AlSb & CO\textsubscript{2} $\rightarrow$ *COOH & 0.12 \\

 & Pristine AlSb & *CHO $\rightarrow$ HCHO & 0.35 \\
HCHO & Fe@AlSb & *OCHO $\rightarrow$ HCOOH & 0.31 \\
 & Co@AlSb & *CHO $\rightarrow$ HCHO & 0.38 \\
 & Ni@AlSb & *CHO $\rightarrow$ HCHO & 0.42 \\

 & Pristine AlSb & *OCHO $\rightarrow$ CH\textsubscript{2}O\textsubscript{2}* & 0.34 \\
CH\textsubscript{3}OH & Fe@AlSb & *OCHO $\rightarrow$ HCOOH & 0.31 \\
 & Co@AlSb & *CHO $\rightarrow$ HCHO & 0.38 \\
 & Ni@AlSb & *OCH\textsubscript{3} $\rightarrow$ CH\textsubscript{3}OH & 0.40 \\

 & Pristine AlSb & *OCHO $\rightarrow$ CH\textsubscript{2}O\textsubscript{2}* & 0.34 \\
CH\textsubscript{4} & Fe@AlSb & *OCHO $\rightarrow$ HCOOH & 0.31 \\
 & Co@AlSb & *CHO $\rightarrow$ HCHO & 0.38 \\
 & Ni@AlSb & HCHO $\rightarrow$ *OCH\textsubscript{3} & 0.28 \\
\bottomrule
\end{tabular}
\end{table}

\subsubsection{Selectivity to HCHO}
{\autoref{pristine11}}(d) shows the pathway to HCHO for pristine AlSb. Here, *CHO to HCHO was the PDS, and overpotential from CO\textsubscript{2} to HCHO conversion for pristine 2D AlSb was 0.35 eV. {\autoref{pristine14}}(d) presents the pathway to HCHO for Fe-doped 2D AlSb. Here, *OCHO to HCOOH was the PDS, and overpotential from CO\textsubscript{2} to HCHO conversion for Fe-doped 2D AlSb was 0.31 eV, which was a significant improvement over that without doping. Fig. S10(d) shows the pathway to HCHO for Co-doped 2D AlSb. Here, *CHO to HCHO was the PDS, and overpotential from CO\textsubscript{2} to HCHO conversion for Co-doped 2D AlSb was 0.38 eV, which showed the negative catalytic effect of doping. We can also see from Fig. S13(d) the pathway to HCHO for Ni-doped 2D AlSb. Here, *CHO to HCHO was the PDS, and overpotential from CO\textsubscript{2} to HCHO conversion for Ni-doped 2D AlSb was 0.42 eV. In this case, doping with Ni acted like an inhibitor. Results are summarized in {\autoref{tbl5}}.  Dopant type was the controlling selection mechanism.

%\clearpage
\begin{figure*}[h]
%\hspace{-2.70cm}
\centering
\includegraphics[scale=0.52]{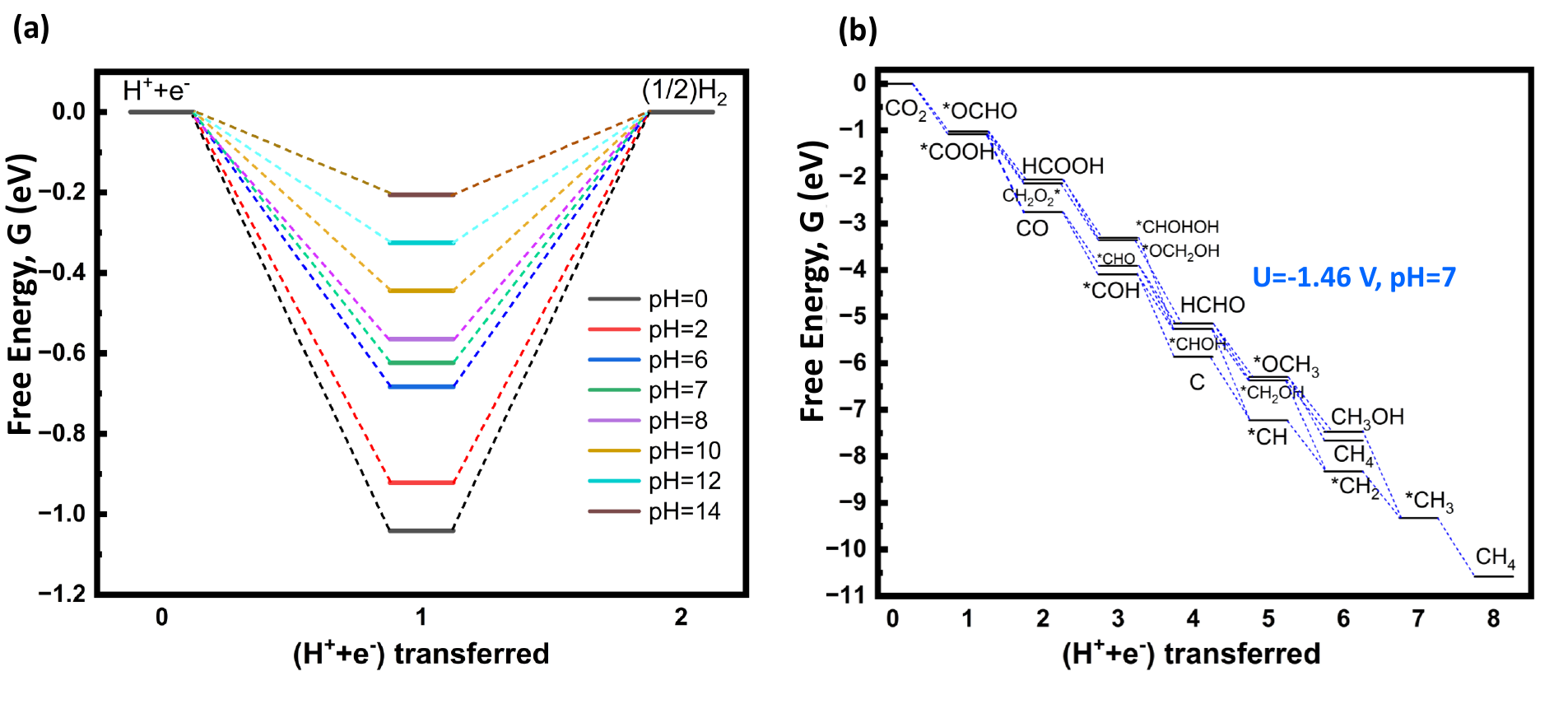}
% \caption{Band structure and density of states of Ni doped(hollow site) AlSb monolayer.}
  \label{pristine24}
\end{figure*}

\begin{figure*}[h]
%\hspace{-2.70cm}
\centering
\includegraphics[scale=0.52]{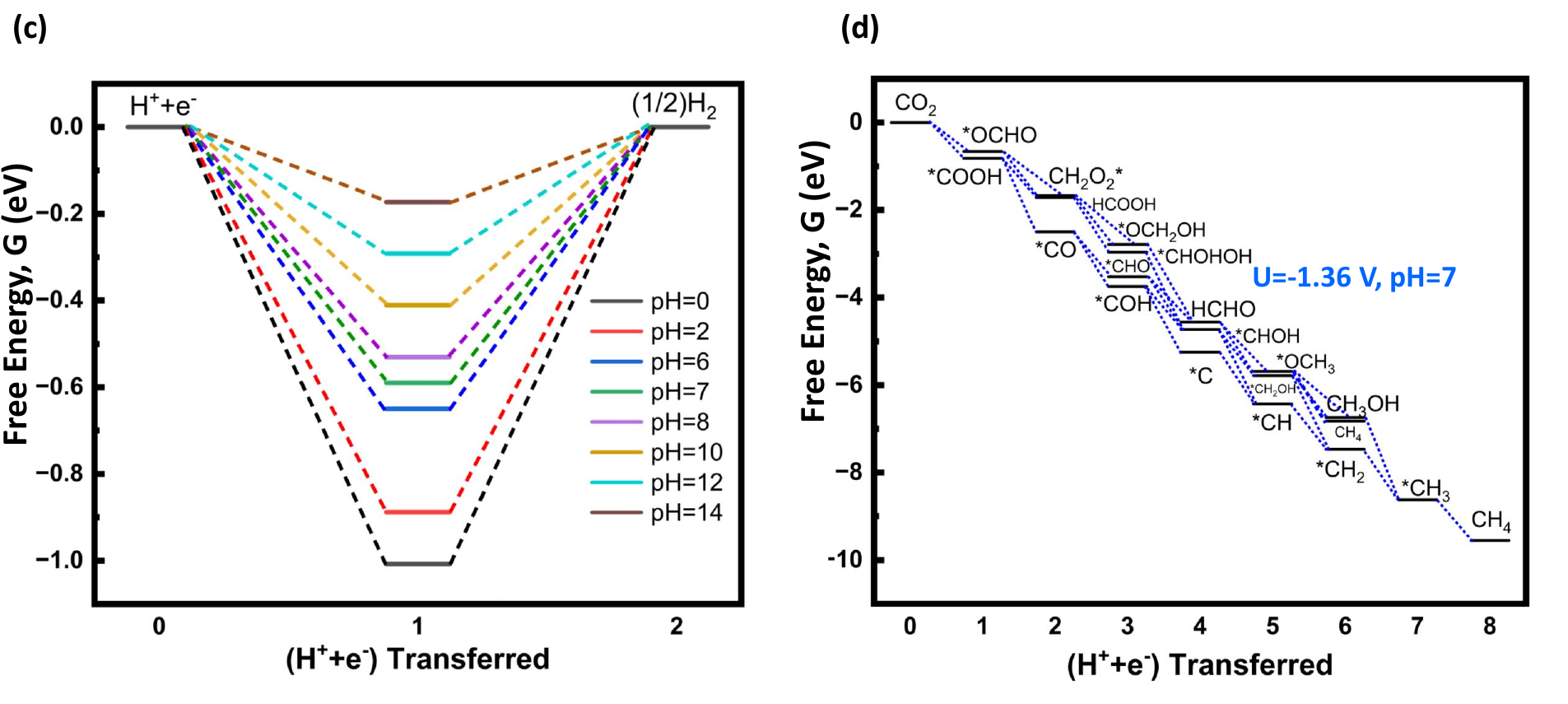}
\caption{Hydrogen evolution reaction (HER) on (a) pristine 2D AlSb and (c) Fe-doped 2D AlSb. Modified pathway under applied potential on (b) pristine 2D AlSb, (d) Fe-doped 2D AlSb.}
  \label{pristine21}
\end{figure*}

\subsubsection{Selectivity to CH$_3$OH}
{\autoref{pristine11}}(e) is showing the pathway to CH\textsubscript{3}OH for pristine AlSb. Results and key findings are summarized in {\autoref{tbl5}}. Here *OCHO to CH\textsubscript{2}O\textsubscript{2}* was the PDS and overpotential from CO\textsubscript{2} to CH\textsubscript{3}OH conversion for pristine 2D AlSb was 0.34 eV. Next {\autoref{pristine14}}.(e) is showing the pathway to CH\textsubscript{3}OH for Fe-doped 2D AlSb. Here, *OCHO to HCOOH was the PDS and overpotential from CO\textsubscript{2} to CH\textsubscript{3}OH conversion for Fe-doped 2D AlSb was 0.31 eV, which was a significant improvement over that of without doping. Figure S11(e) shows the pathway to CH\textsubscript{3}OH for Co-doped 2D AlSb. Here, *CHO to HCHO was the PDS and overpotential from CO\textsubscript{2} to CH\textsubscript{3}OH conversion for Co-doped 2D AlSb was 0.38 eV, which showed the negative catalytic effect of doping with Co. And finally, Fig. S14(e) shows the pathway to CH\textsubscript{3}OH for Ni-doped 2D AlSb. Here, *OCH\textsubscript{3} to CH\textsubscript{3}OH was the PDS and overpotential from CO\textsubscript{2} to CH\textsubscript{3}OH conversion for Ni-doped 2D AlSb was 0.40 eV. In this case, doping with Ni was acting like a negative catalyst.

\begin{figure*}[h]
%\hspace{-2.70cm}
\centering
\includegraphics[scale=0.52]{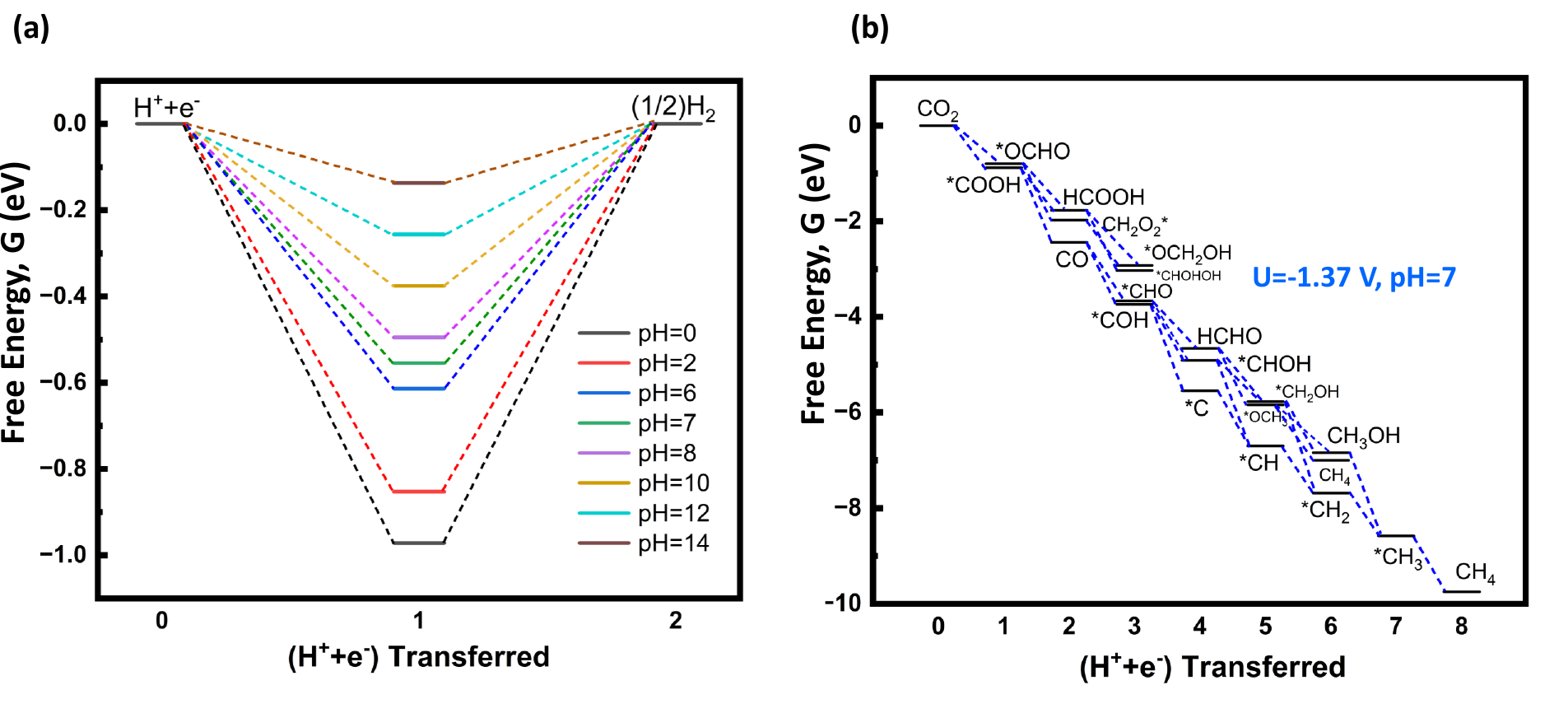}
% \caption{Band structure and density of states of Ni doped(hollow site) AlSb monolayer.}
  \label{pristine22}
\end{figure*}

%\clearpage
\begin{figure*}[H]
%\hspace{-2.70cm}
\centering
\includegraphics[scale=0.52]{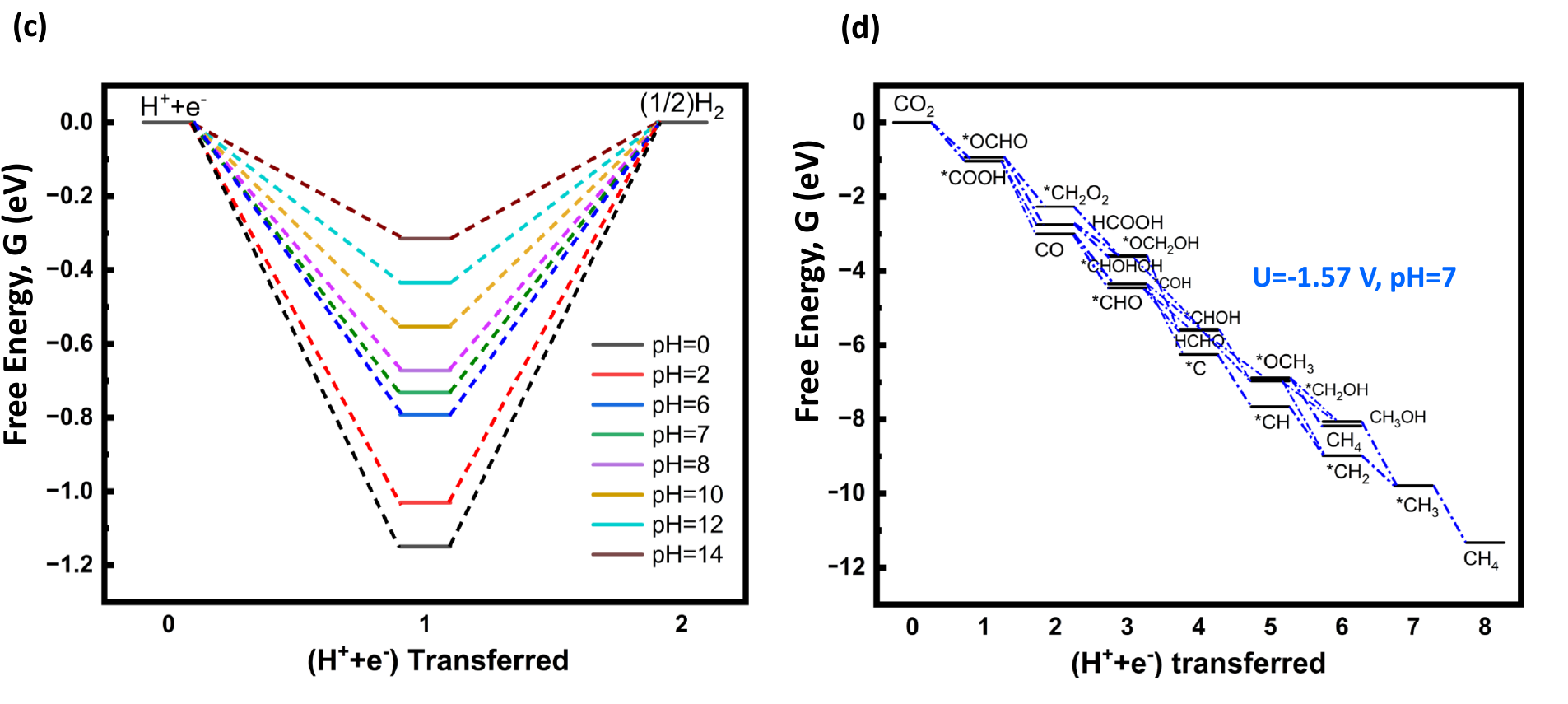}
\caption{Hydrogen evolution reaction (HER) on (a) Co-doped 2D AlSb and (c) Ni-doped 2D AlSb. Modified pathway under applied potential on (b) Co-doped 2D AlSb and (d) Ni-doped 2D AlSb.}
  \label{pristine23}
\end{figure*}

\subsubsection{Selectivity to CH$_4$}
{\autoref{pristine11}}(f) is showing the pathway to CH\textsubscript{4} for pristine AlSb. Here, all the key findings are summarized in {\autoref{tbl5}}. We can see in \autoref{pristine11} that, *OCHO to CH\textsubscript{2}O\textsubscript{2}* was the PDS and overpotential from CO\textsubscript{2} to CH\textsubscript{4} conversion for pristine 2D AlSb was 0.34 eV. The following {\autoref{pristine14}}(f) is showing the pathway to CH\textsubscript{4} for Fe-doped 2D AlSb. Here, *OCHO to HCOOH was the PDS and overpotential from CO\textsubscript{2} to CH\textsubscript{4} conversion for Fe-doped 2D AlSb was 0.31 eV, which was a significant improvement over that of without doping. Next, Fig. S11(f) shows the pathway to CH\textsubscript{4} for Co-doped 2D AlSb. Here, *CHO to HCHO was the PDS and overpotential from CO\textsubscript{2} to CH\textsubscript{4} conversion for Co-doped 2D AlSb was 0.38 eV. And finally, Fig. S14(f) shows the pathway to CH\textsubscript{4} for Ni-doped 2D AlSb. Here, HCHO to *OCH\textsubscript{3} was the PDS, and overpotential from CO\textsubscript{2} to CH\textsubscript{4} conversion for Ni-doped 2D AlSb was 0.28 eV, which was an extremely fine catalytic effect. In this case, doping with Ni acted like a positive catalyst. By discussing all four cases, a conclusion can be drawn that both doping and applied electric potential can be two crucial factors to select a specific CO\textsubscript{2}RR product. Also, a specific product could be reached with minimum overpotential if the dopant were varied. On the other hand, if the catalyst (dopant) was fixed, then different applied potentials could lead to different useful reduction products.

\begin{table}[h]%[width=.9\linewidth,cols=4,pos=h]
\centering
\caption{HER barrier and applied potentials for different catalysts}\label{tbl9}
\begin{tabular}{ccc}
\toprule
\makecell{\textbf{Catalyst}} & \makecell{\textbf{Barrier of HER} \\ \textbf{(eV)}}  & 
\makecell{\textbf{Applied Voltage} \\ \textbf{(V)}}
\\
\midrule
Pristine AlSb & -1.04 & -1.46 \\
Fe@AlSb & -1.01 & -1.36 \\
Co@AlSb & -0.97 & -1.37\\
Ni@AlSb & -1.15 & -1.57 \\
\bottomrule
\end{tabular}
\end{table}
%\clearpage

\subsubsection{HER and effects of pH}
H\textsubscript{2} ion (H\textsuperscript{+}) from aqueous solution, or vapor, or humidity can take part in a reduction reaction: 2H\textsuperscript{+}+2e\textsuperscript{-}$\rightarrow$ H\textsubscript{2}. This reaction is called HER. It can compete with CO\textsubscript{2}RR if the overpotential of HER is less than that of CO\textsubscript{2}RR. That's why HER was considered to be an unwanted side reaction. HER of pristine and Fe-doped AlSb are shown in \autoref{pristine21}(a),(c). HER of Co and Ni-doped AlSb are shown in {\autoref{pristine23}}(a),(c). We can see from these figures that the energy barrier, even at the highest pH (pH=14, where [H\textsuperscript{+}] was lowest in concentration), was less than that of CO\textsubscript{2}RR. Therefore, the application of potential was required to make this side reaction less likely.

\subsubsection{Modified pathway with applied potential}
For pristine AlSb, {\autoref{pristine23}}(a) shows that \( \Delta G_{\text{HER}} = -1.04 \) eV where \(\Delta G_{\text{Max}} = 0.42\) eV for pathway to HCOOH. That's why a voltage (electric potential) needed to be applied so that HER became less likely. The minimum amount of potential came from the condition at pH = 7 (neutral medium), \(\Delta G_{\text{HER}} > \Delta G_{\text{Max,old}} - neU\). These values are tabularized in \textcolor{black}{\autoref{tbl9}}. The modified pathways after applying voltage and pH are shown in  \autoref{pristine21}(b),(d) and \autoref{pristine23}(b),(d). Calculated Gibbs free energy under the applied potential is shown in Figs. S15-S17.

%\clearpage
\begin{figure*}[h]
%\hspace{-2.70cm}
\centering
\includegraphics[scale=0.52]{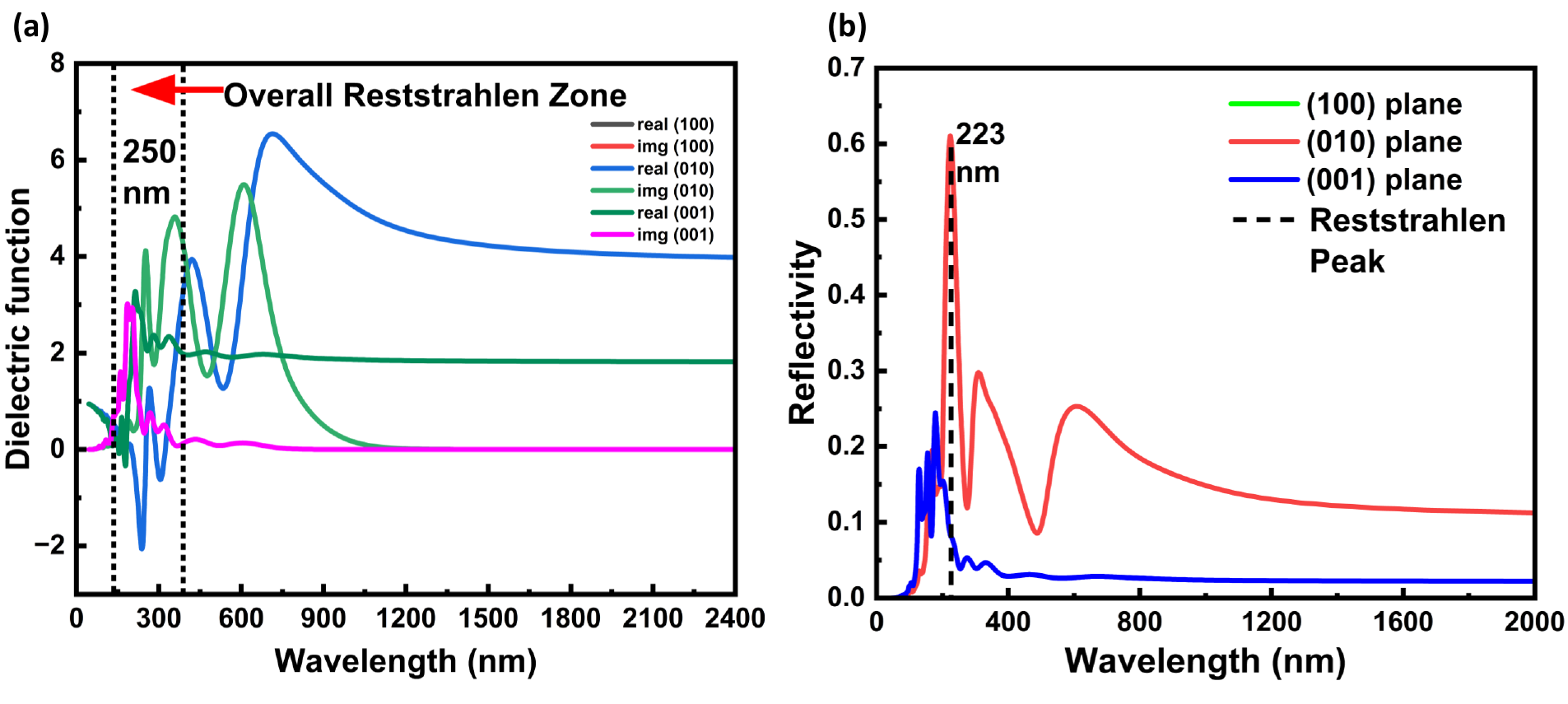}
%\caption{Hydrogen evolution reaction (HER) on (a) pristine AlSb, (c) Fe doped, (e) Co doped (g) ,Ni doped AlSb and modified pathway under applied potential on (b) pristine AlSb, (d) Fe doped, (f) Co doped, (h) Ni doped AlSb respectively.}
  \label{pristine27}
\end{figure*}

\begin{figure*}[h]
%\hspace{-2.70cm}
\centering
\includegraphics[scale=0.52]{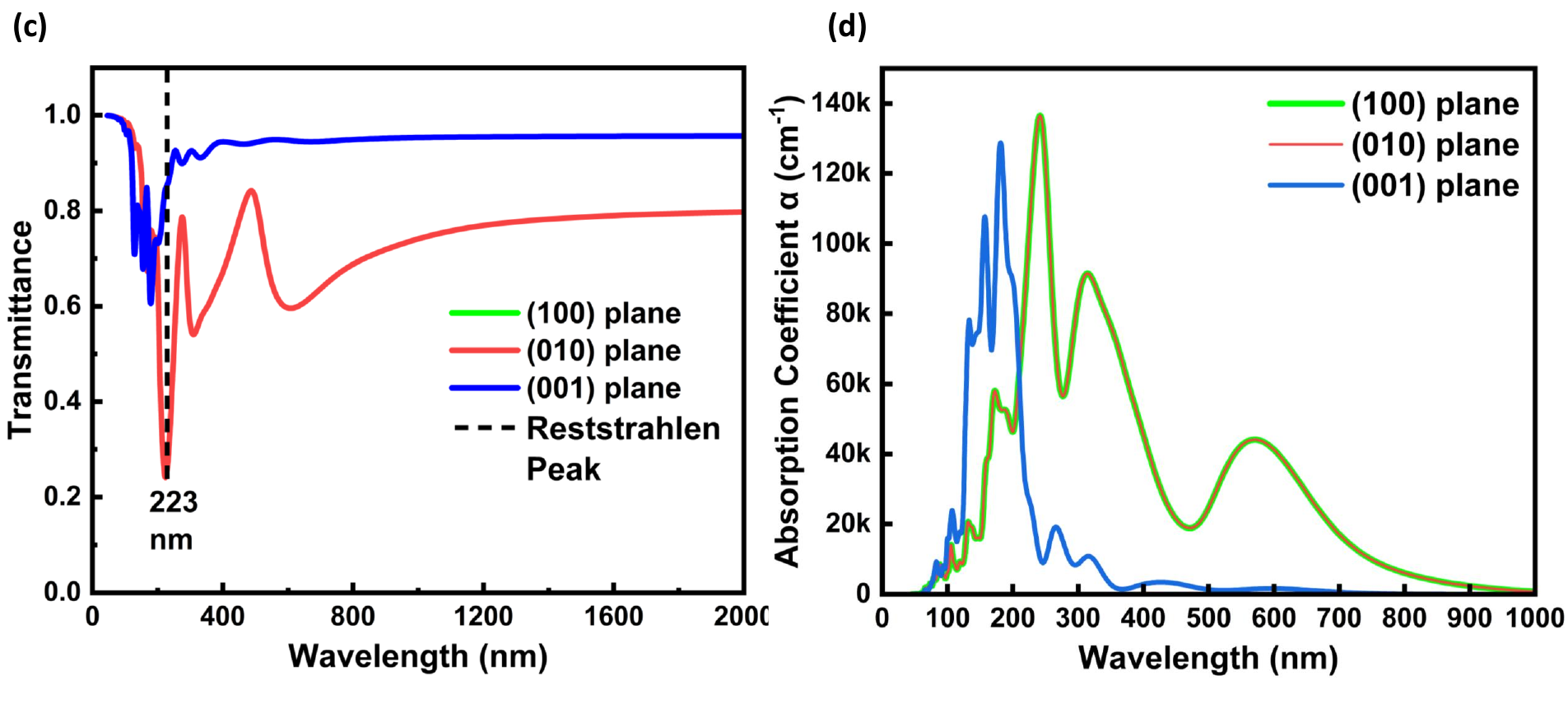}
\caption{(a) Dielectric function vs wavelength (both real and imaginary part) along with reststrahlen zone, (b) reflectance vs wavelength with reststrahlen peak, (c) transmittance vs wavelength with reststrahlen peak, (d) absorption coefficient vs wavelength.}
  \label{pristine28}
\end{figure*}

%\clearpage

\subsection{Optical properties}
Four fundamental optical properties: dielectric constant (real and imaginary part), reflectivity, transmittance, and absorption coefficient are graphically exhibited in {\autoref{pristine28}}(a), (b), (c), and
(d), respectively in three different planes (100), (010), and (001). All the optical properties in (100) and (010) planes were identical, and therefore these two graphs superimposed on each other. In both (100) and (010) planes, a significant property reststrahlen effect \cite{ribbing1991reststrahlen,adachi1999reststrahlen} was observed where a semiconductor behaved like a metal with high reflectivity. Reststrahlen zone is the range of wavelength where the real part of the dielectric constant becomes negative, and the imaginary part of the dielectric constant is very high. In this region, material behaves like a metal/reflector \cite{feng2015photonic}. The pristine 2D AlSb showed the reststrahlen effect in the UV region with an interval of
250 nm (reststrahlen zone), and the reststrahlen peak was located at the maximum reflectance point $\lambda$=223 nm in (100) and (010) plane. Hence, one application of 2D AlSb can be as an UV filter with $\lambda$=223 nm in (100) and (010) planes. Transmittance in \textcolor{black}{ \autoref{pristine28}}(c) was calculated by,
\begin{equation}
T_{\Sigma} = \frac{(1 - R)^2 \exp(-\alpha W)}{1 - R^2 \exp(-2\alpha W)},
    \label{eq7}
\end{equation}

where $\alpha$ and R were from the simulation, and W was the same as buckling height h. Similarly, absorbance(A) \cite{wong2015method} was also calculated following the simple formula:\begin{equation}
       A=1-R-T
   \label{eq8}
\end{equation}
{ \autoref{pristine28}}(d) is showing the absorption spectra. The absorption spectra in the (100) plane and (010) were also identical, and the absorption spectra were different in the (001)
plane. The absorption spectra were broad enough in the (100) and (010) plane to cover the entire
visible wavelength range, and peaked at the UV or violet region. Another potential application of AlSb can be a solar cell as well as a UV or violet photo-detector. Although the value of the absorption coefficient $\alpha$ was very high, the thickness of monolayer W was in the {\AA} range. 

%That's why absorbance A (percentage of absorption) was not very high.

\begin{table}[h]
\scriptsize
\centering
\caption{Maximum Gibbs free energy changes ($\Delta G$) in eV for various CO$_2$RR catalysts and intermediate complexes, comparing with literature}
\label{tbl:gibbs}
\begin{tabular}{p{1.2cm}cccccp{0.35cm}}
\hline
\textbf{Catalyst} & \textbf{CO} & \textbf{HCOOH} & \textbf{HCHO} & \textbf{CH$_3$OH} & \textbf{CH$_4$} & \textbf{Ref.} \\
\hline
FeN\textsubscript{4}@Gr & 0.60eV & 0.72eV & 0.87eV & 0.87eV & 0.87eV & \cite{yang2022theoretical} \\
% other rows..
CoN\textsubscript{4}@Gr & 0.68eV & 0.85eV & 0.68eV & 0.68eV & 0.68eV & \cite{yang2022theoretical} \\
NiN\textsubscript{4}@Gr & 1.70eV & 1.51eV & 1.70eV & 1.70eV & 1.70eV & \cite{yang2022theoretical} \\
CrN-PDN    & -        & 0.38eV  & -        & -        & -        &  {\textcolor{black}{\cite{yang2023theoretical}}}\\
CrN-PON    & -        & 0.49eV  & -        & -        & -        &  {\textcolor{black}{\cite{yang2023theoretical}}}\\
MnN\textsubscript{4}    & -        & 0.33eV  & -        & -        & -        &  {\textcolor{black}{\cite{yang2023theoretical}}}\\
MnN\textsubscript{4}-GN   & -        & 0.32eV  & -        & -        & -        &  {\textcolor{black}{\cite{yang2023theoretical}}}\\

MnN\textsubscript{4}-PDN   & -        & 0.23eV  & -        & -        & -        &  {\textcolor{black}{\cite{yang2023theoretical}}}\\
% other rows...

MnN\textsubscript{4}-PON   & -        & 0.20eV  & -        & -        & -        &  {\textcolor{black}{\cite{yang2023theoretical}}}\\
FeN\textsubscript{4}-GN   & -        & 0.35eV  & -        & -        & -        &  {\textcolor{black}{\cite{yang2023theoretical}}}\\
FeN\textsubscript{4}-PDN   & -        & 0.55eV  & -        & -        & -        &  {\textcolor{black}{\cite{yang2023theoretical}}}\\
RhN\textsubscript{4}-GN   & -        & 0.39eV  & -        & -        & -        &  {\textcolor{black}{\cite{yang2023theoretical}}}\\
CoN\textsubscript{4}-PDN   & -        & -  & 0.55eV        & -        & -        &  {\textcolor{black}{\cite{yang2023theoretical}}}\\
CoN\textsubscript{4}-PON   & -        & -  & 0.60eV        & -        & -        &  {\textcolor{black}{\cite{yang2023theoretical}}}\\

Ti@Ti\textsubscript{3}C\textsubscript{2}O\textsubscript{2} & 2.50eV  & 3.00eV  & -        & -        & 0.99eV  &  {\textcolor{black}{\cite{zhou2023efficient}}}\\
Co@Ti\textsubscript{3}C\textsubscript{2}O\textsubscript{2} & 1.70eV  & 1.50eV  & -        & -        & 0.21eV  &  {\textcolor{black}{\cite{zhou2023efficient}}}\\
Ni@Ti\textsubscript{3}C\textsubscript{2}O\textsubscript{2} & 1.49eV  & 0.85eV  & -        & -        & -  &  {\textcolor{black}{\cite{zhou2023efficient}}}\\
Sc@Ti\textsubscript{3}C\textsubscript{2}O\textsubscript{2} & -  & -  & -        & -        & 0.49eV &  {\textcolor{black}{\cite{zhou2023efficient}}}\\

Co@MoS\textsubscript{2}    & -        & -        & -        & -        & 0.24eV  &  {\textcolor{black}{\cite{ren2022single}}}\\
Ru@Cu(211)    & -        & -        & -        & -        & 0.44eV  &  {\textcolor{black}{\cite{ren2022single}}}\\
Fe@MoS\textsubscript{2}    & -        & -        & -        & -        & 0.39eV  &  {\textcolor{black}{\cite{ren2022single}}}\\

Bi\textsubscript{2}WO\textsubscript{6}     & 1.90eV  & -        & -        & -        & -        &  {\textcolor{black}{\cite{liu2022effect}}}\\
Cu(211)     & -  & -        & 0.91eV        & 0.90eV        & -        &  {\textcolor{black}{\cite{liu2022effect}}}\\

Cu-Doped Phosphorene & -  & -        & 0.61 eV  & 0.60 eV  & -        &  {\textcolor{black}{\cite{zhang2021co}}}\\

AlSb  & -0.20eV & 0.42eV  & 0.35eV  & 0.34eV  & 0.34eV  &  \\
Fe@AlSb & -0.12eV & 0.31eV & 0.31eV & 0.31eV & 0.31eV &  \\
Co@AlSb & -0.12eV & 0.40eV & 0.38eV & 0.38eV & 0.38eV &  \\
Ni@AlSb & 0.12eV & 0.12eV & 0.42eV & 0.40eV & 0.28eV &  \\
\hline

\end{tabular}
\end{table}

\subsection{Comparative analysis}
In \autoref{tbl:gibbs}, a list of previously analyzed and simulated CO\textsubscript{2}RR catalysts, along with their lowest overpotentials for different intermediate complexes of CO\textsubscript{2}RR are tabularized with appropriate citations. The last four rows are the experimental results of this paper. From \autoref{tbl:gibbs}, we can observe that graphene derivatives (GN, PDN, PON) required very high overpotential in most of the catalysts. Ti@Ti\textsubscript{3}C\textsubscript{2}O\textsubscript{2} and Bi\textsubscript{2}WO\textsubscript{6} were hard to synthesize and needed very high overpotential as well. Cu(111) and Cu-doped phosphorene were easily available; however, their overpotential was moderately high. Although Co@MoS\textsubscript{2} had a lower overpotential, it had toxicity and some synthesis problems. Hence, it can be concluded from \autoref{tbl:gibbs} that there was hardly any catalyst with a lower overpotential synthesized or simulated before than AlSb and doped AlSb for CO$_2$RR. Moreover, 2D AlSb was easily fabricable \cite{kim2024future}. This
research clearly proved that AlSb was spectacular and extraordinary for decreasing the CO$_2$RR barrier and enhancing favorability. The reasons were the large surface area, atomic thinness of 2D materials, excellent mass transport, reducing diffusion limitations, and improving reactant accessibility in the case of 2D AlSb. The atomic dispersion of metal atoms on 2D AlSb maximized active site utilization, leading to higher catalytic efficiency with minimal material usage in the case of doped AlSb. 
%\clearpage

\section{Conclusion}
While CO\textsubscript{2} is essential for plant growth, beverages, and various industrial applications, its excessive atmospheric concentration, surpassing the optimum level, results in global warming, ocean acidification, and adverse health impacts. We explored 2D AlSb and its derivatives as CO\textsubscript{2}RR catalysts and performed necessary simulations to mitigate the effects of CO\textsubscript{2}. We demonstrated 2D AlSb to be an excellent CO\textsubscript{2}RR catalyst, exhibiting low overpotential and high efficiency in both pristine and doped forms. 2D AlSb also showed remarkable product selectivity, with the lowest overpotentials obtained for HCOOH on Ni-doped AlSb (0.12 eV), HCHO on Fe-doped AlSb (0.31 eV), CH\textsubscript{3}OH on pristine AlSb (0.31 eV), and CH\textsubscript{4} on Ni-doped AlSb (0.28 eV). We found Fe-doped 2D AlSb to be the best CO\textsubscript{2}RR catalyst. Because Fe-doped 2D AlSb exhibited the least bandgap and the most significant 3d orbital contribution in PDOS. We ensured the contribution of 3d orbital from the projected band structure as well. Also, Fe@AlSb required the lowest voltage for HER mitigation and comparatively lower overpotential. Negative binding energies confirmed AlSb and its derivatives to be energetically stable. We proved AlSb to be a highly stable material as a substrate via phonon dispersion and AIMD simulations. We applied electric potentials: -1.46 V for pristine AlSb, -1.36 V for Fe@AlSb, -1.37 V for Co@AlSb, and -1.57 V for Ni@AlSb to mitigate the effects of HER. Controlling factors of selectivity were pH, potential, and dopant. Doping with Fe, Co, and Ni increased the catalytic effect of 2D AlSb due to bandgap reduction and 3d orbital contribution, as explained in band structure, projected band structure plots, DOS, and PDOS. From CDD and Bader calculations, we explicitly proved charge transfer from adsorbate to substrate both qualitatively and quantitatively. 2D AlSb and its derivatives are highly efficient and reactive nanomaterial interfaces for CO\textsubscript{2}RR adsorbates, which will revolutionize eco-friendly, sustainable CO\textsubscript{2} consumption in industry, keeping the environment secure.

%The author names and affiliations %could be formatted in two ways:
%\begin{enumerate}[(1)]
%\item Group the authors per affiliation.
%\item Use footnotes to indicate the affiliations.
%\end{enumerate}
%See the front matter of this document for examples. 
%You are recommended to conform your choice to the journal you 
%are submitting to.

%\section{Floats}
%{Figures} may be included using the command,\linebreak 
%\verb+\includegraphics+ in
%combination with or without its several options to further control
%graphic. \verb+\includegraphics+ is provided by {graphic[s,x].sty}
%which is part of any standard \LaTeX{} distribution.
%{graphicx.sty} is loaded by default. \LaTeX{} accepts figures in
%the postscript format while pdf\LaTeX{} accepts {*.pdf},
%{*.mps} (metapost), {*.jpg} and {*.png} formats. 
%pdf\LaTeX{} does not accept graphic files in the postscript format. 

%\begin{figure}
%	\centering
%		\includegraphics[scale=.75]{figs/Fig1.pdf}
%	\caption{The evanescent light - $1S$ quadrupole coupling
%	($g_{1,l}$) scaled to the bulk exciton-photon coupling
%	($g_{1,2}$). The size parameter $kr_{0}$ is denoted as $x$ %and
%	the \PMS is placed directly on the cuprous oxide sample %%%%($\delta
%	r=0$, See also Table \protect\ref{tbl1}).}
%	\label{FIG:1}
%\end{figure}

%The \verb+table+ environment is handy for marking up tabular
%material. If users want to use {multirow.sty},
%{array.sty}, etc., to fine control/enhance the tables, they
%are welcome to load 

%any package of their choice and
%{cas-dc.cls} will work in combination with all loaded
%packages.

\printcredits
\section*{Declaration of competing interest}
The authors declare that they have no competing financial interests or personal relationships that could have influenced this work.

\section*{Data availability statement}
{The data that support the findings of this study are not publicly available but can be obtained from the authors upon reasonable request.} 

\section*{Acknowledgements}
M.M.I. and A.Z. gratefully acknowledge the Bangladesh University of Engineering and Technology (BUET) and the Bangladesh Research and Education Network (BdREN) for providing computational resources and technical support.

%\section{Supplementary material}
%Supplementary material includes Figs. (S1–S17) showing additional information for CO\textsubscript{2}RR with Co-doped 2D AlSb and Ni-doped 2D AlSb structures. It can be found online at https://.... 

%\bibliographystyle{cas-model2-names}

\bibliography{cas-refs}
\balance


%\clearpage
\appendix
\appendixpage
\section{Structural properties}

\begin{figure*}[h]
%\hspace{-2.70cm}
\centering
\includegraphics[scale=0.38]{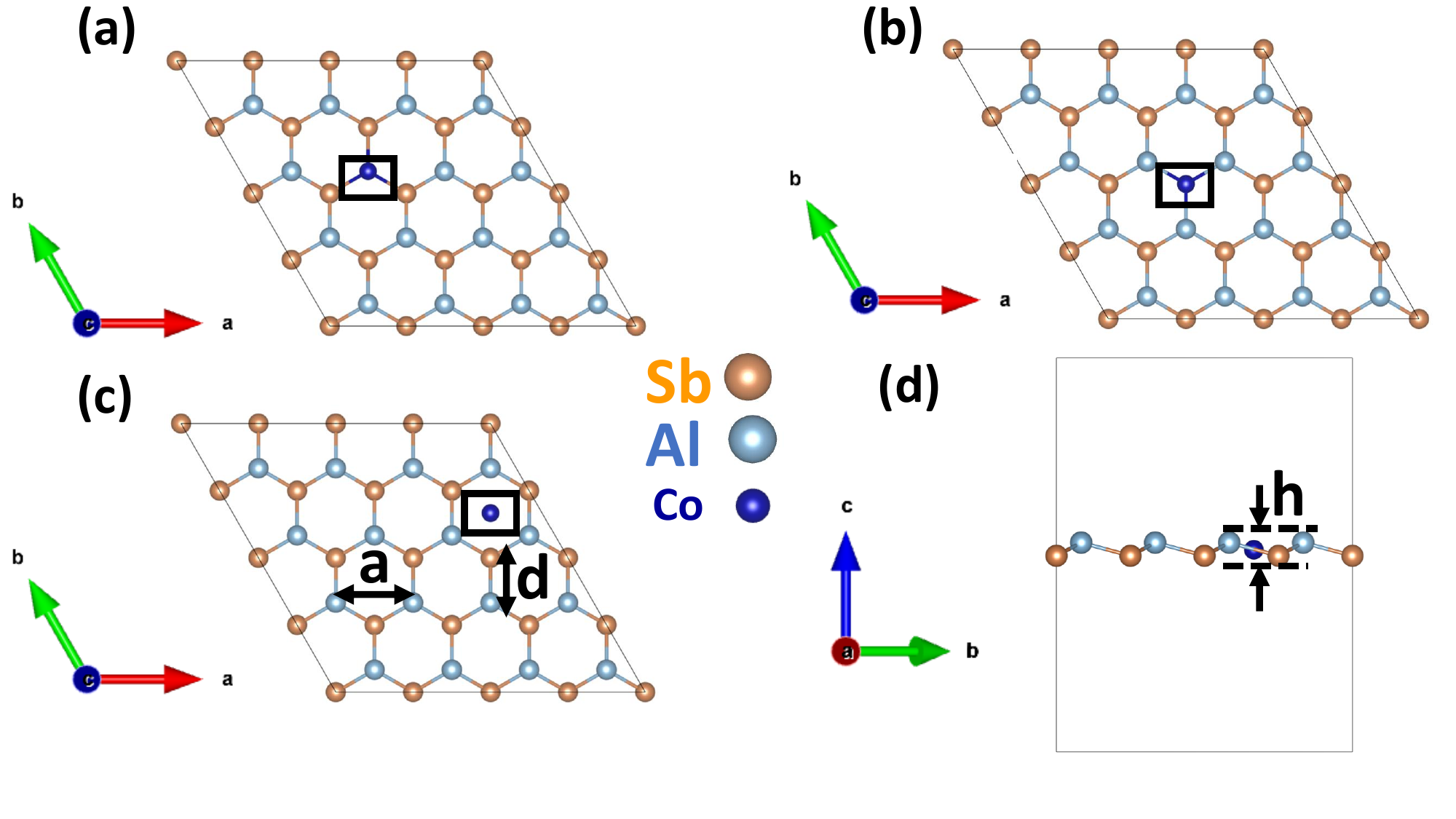}
  \caption{Schematics of Co-doped 2D AlSb. Co replacing (a)top site, (b)bottom site and (c)hollow site. (d)Side view of hollow site. Here, a is the lattice constant, d is the bond length and h is the buckling height.}
  \label{pristine331}
\end{figure*}

\begin{figure*}[h]
%\hspace{-2.70cm}
\centering
\includegraphics[scale=0.38]{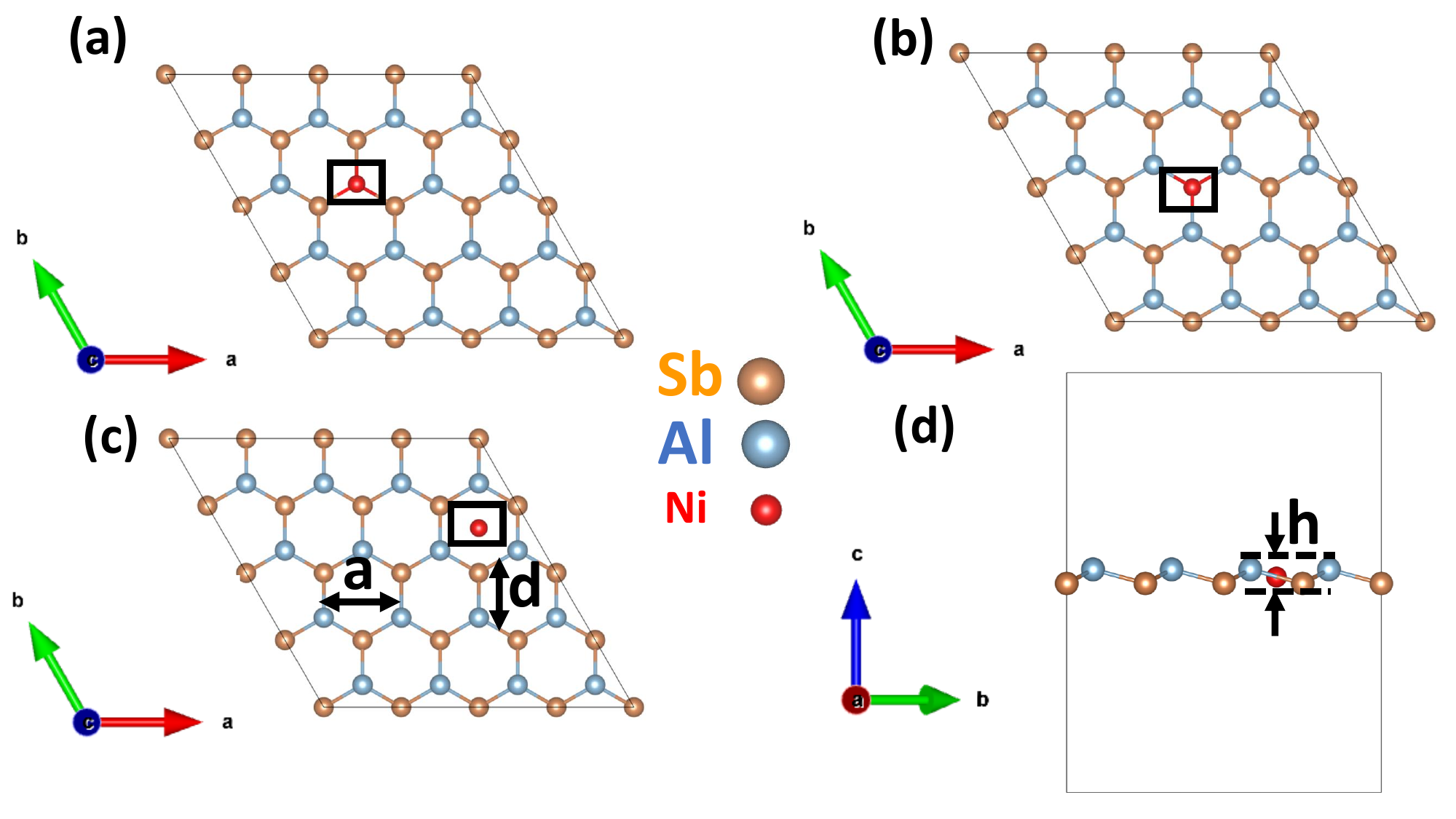}
  \caption{Schematics of Ni-doped 2D AlSb. Ni replacing (a)top site, (b)bottom site and (c)hollow site. (d)Side view of hollow site. Here, a is the lattice constant, d is the bond length and h is the buckling height.}
  \label{pristine431}
\end{figure*}

The unit cell of hexagonal AlSb was created by cleaving (111) surface of zinc blende AlSb structure in Cambridge Serial Total Energy Package (CASTEP). The unit cell of hexagonal AlSb was optimized. The lattice constant a, bond length d and buckling height h were found to be 4.3986 \AA, 2.6190 {\AA} and 0.6400 {\AA} 
respectively from optimized structure %{\textcolor{blue}{\cite{shahriar2023halogenation}}}
indicated in \autoref{pristine331}-\autoref{pristine431}.  
 Then a (4x4x1) supercell was created as a monolayer of AlSb which worked as a substrate for the dopants (Fe, Co, Ni) and  an adsorption interface for CO\textsubscript{2} reduction reaction (CO\textsubscript{2}RR) intermediate complexes.  There were basically three types of active sites for any foreign atom to be occupied. (i) Top site (Al atom will be replaced), (ii) bottom site (Sb atom will be replaced), and (iii) hollow site (interatomic space will be occupied). Both pristine and doped AlSb were checked as CO\textsubscript{2}RR electrocatalysts. The Co- and Ni-doped figures are explicitly shown in \autoref{pristine331}-\autoref{pristine431}. Among the three sites, the hollow site was the most energetically favorable structure for Fe, Co, and Ni. The hollow site was chosen for further calculations such as band structures, density of states, orbital projected band structures and density of states, Gibbs free energy, and HER.

%\clearpage

%\clearpage

\clearpage
\section{Structures of intermediate complexes}
\begin{figure*}[h]
%\hspace{-2.70cm}
\centering
\includegraphics[scale=0.36]
{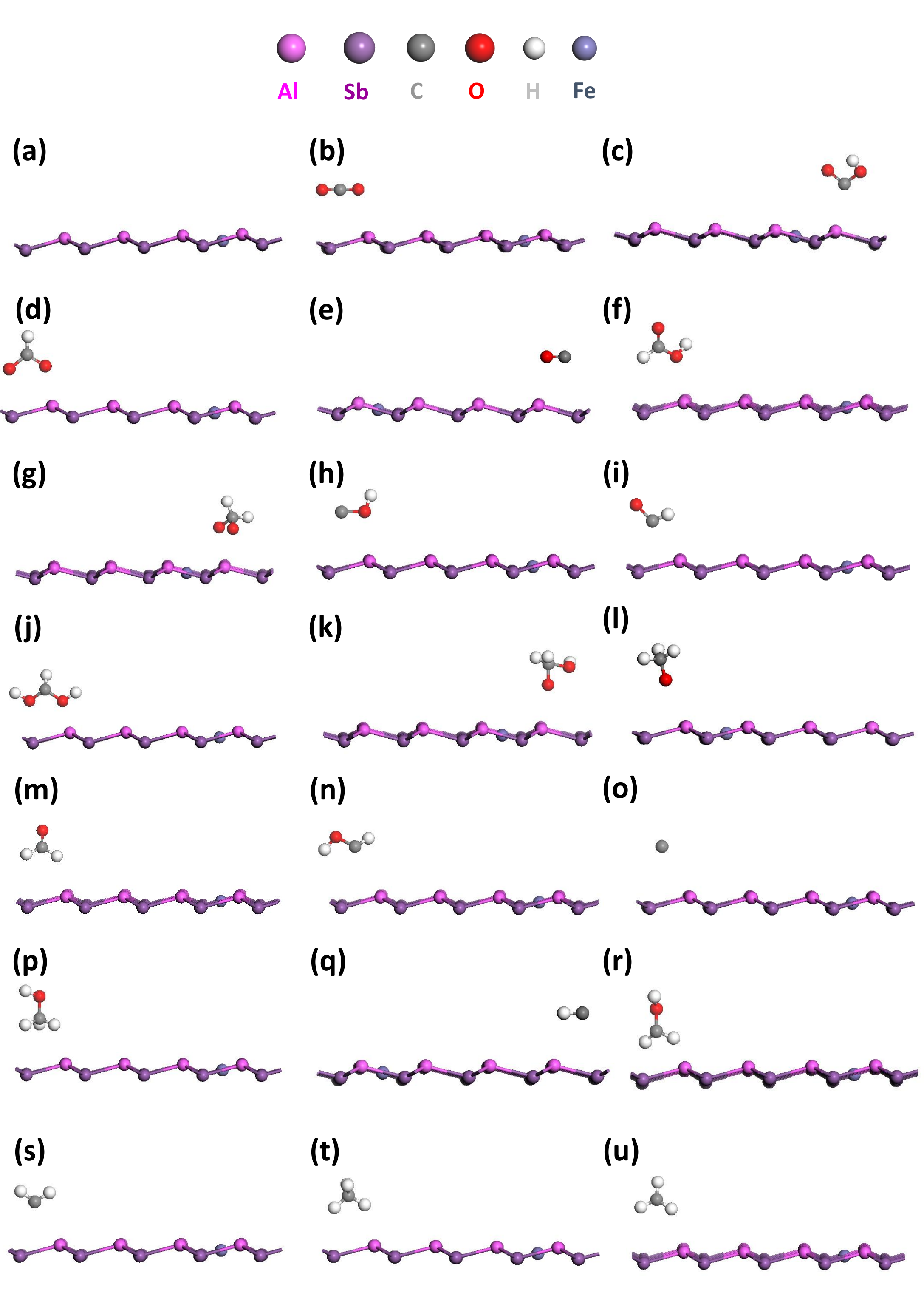}
  \caption{Schematics of (a)Fe-doped 2D AlSb and schematics of (b)CO\textsubscript{2}, (c)*COOH, (d)*OCHO, (e)CO, (f)HCOOH, (g)*CH\textsubscript{2}O\textsubscript{2}, (h)*COH, (i)*CHO, (j)*CHOHOH, (k)*OCH\textsubscript{2}OH, (l)*OCH\textsubscript{3}, (m)HCHO, (n)*CHOH, (o)*C, (p)CH\textsubscript{3}OH, (q)*CH, (r)*CH\textsubscript{2}OH, (s)*CH\textsubscript{2}, (t)CH\textsubscript{4}, (u)*CH\textsubscript{3} adsorbed on Fe-doped 2D AlSb respectively.}
  \label{pristine400}
\end{figure*}

We've found structures of 21 intermediate complexes required for CO\textsubscript{2}RR in eight electronic steps. All the intermediate complexes necessary for the entire pathway of Fe-doped AlSb are shown in \autoref{pristine400}. Gibbs free energy calculations were performed on each of intermediate complexes adsorbed on pristine 2D AlSb, Fe-doped 2D AlSb, Co-doped 2D AlSb and Ni-doped 2D AlSb.

21 intermediate complexes were considered for further calculations of Gibbs free energy to determine the overpotential and reaction favorability of CO\textsubscript{2}RR. The results are summarized in \autoref{tbl:gibbs}. The results are summarized in \autoref{tbl:gibbs}. The results are summarized in \autoref{tbl:gibbs}. The results are summarized in \autoref{tbl:gibbs}. The results are summarized in \autoref{tbl:gibbs}. The results are summarized in \autoref{tbl:gibbs}.
%\clearpage

\section{Electronic properties of Co- and Ni-doped 2D AlSb}
\begin{figure*}[h]
%\hspace{-2.70cm}
\centering
\includegraphics[width=\textwidth]{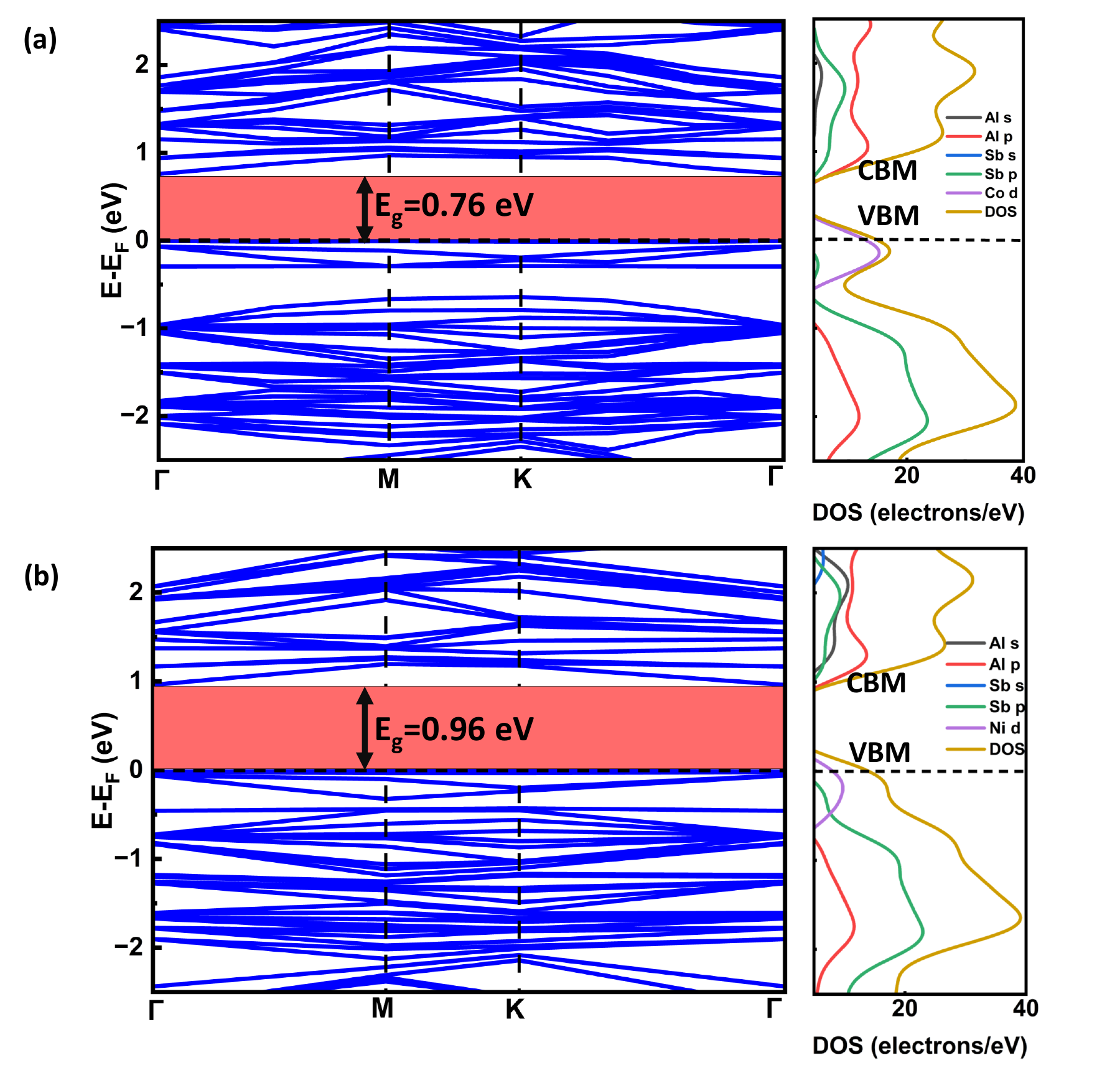}
\caption{Band structure, density of states and projected density of states of (a) Co-doped (hollow site) 2D AlSb and (b) Ni-doped (hollow site) 2D AlSb.}
  \label{pristine511}
\end{figure*}

Three fundamental electronic properties, band structure, density of states (DOS) and projected density of states (PDOS) of Co- and Ni-doped 2D AlSb are indicated in \autoref{pristine511}. By observing the band structure, DOS and PDOS from \autoref{pristine511}, we can notice peaks around Fermi level. Co-doped 2D AlSb showed a medium peak and Ni-doped 2D AlSb showed the lowest peak. Actually, this figure indicated that the 3d orbital electronic effect of Co was moderate and the 3d orbital electronic effect of Ni was minimum. So, the reason for a gradual bandgap reduction for Co- and Ni-doped catalysts was 3d orbital contribution from Co and Ni.

\section{Orbital projected band structure of Fe-doped 2D AlSb}
\begin{figure*}[h]
%\hspace{-2.70cm}
\centering
\includegraphics[width=\textwidth]{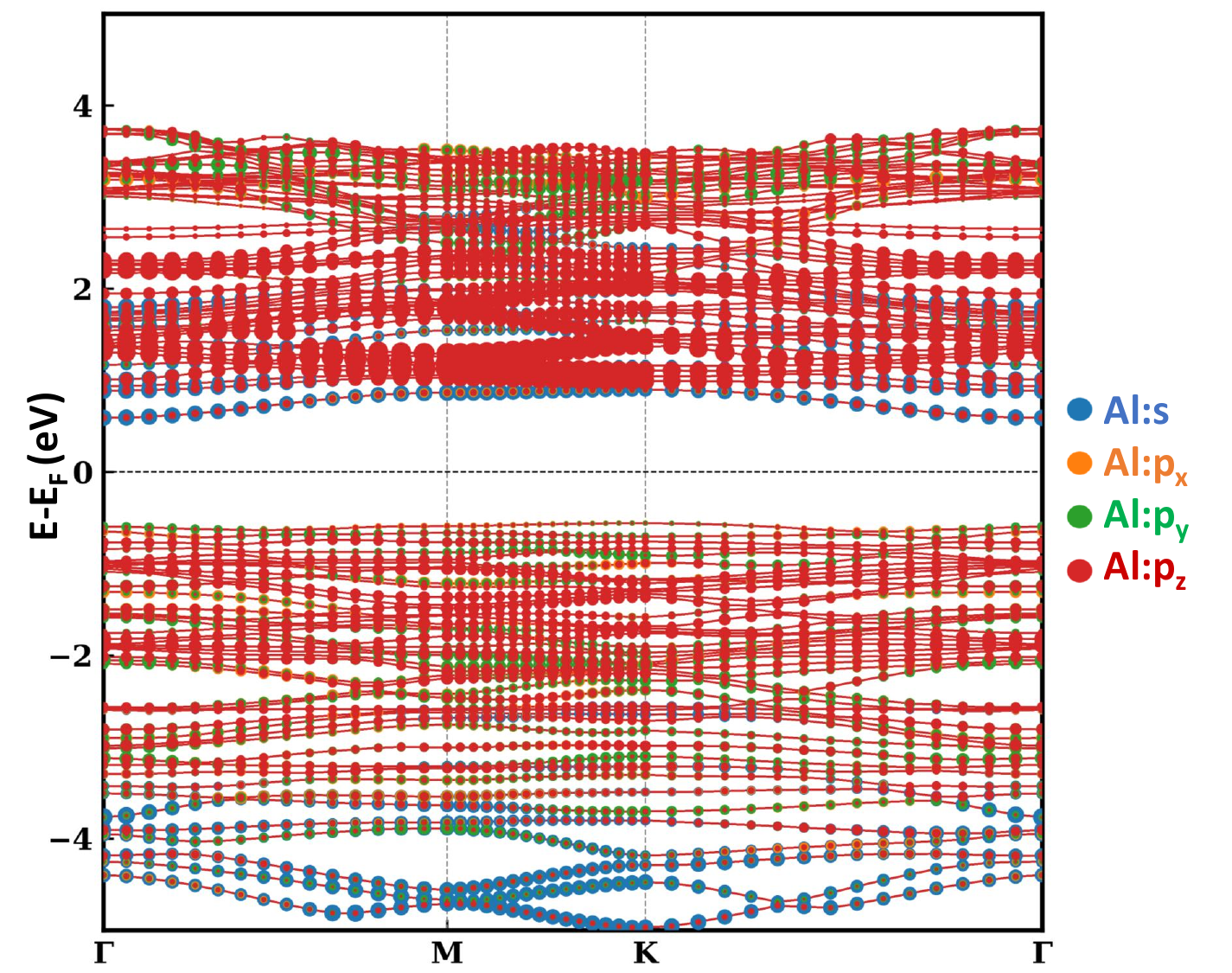}
  \caption{Orbital projected band structure of Fe-doped 2D AlSb, highlighting the contributions from Al-s and Al-p orbitals along the high-symmetry k-path. The circle size represents the orbital weight.}
  \label{pristine4201}
\end{figure*}

\autoref{pristine4201} is showing the orbital projected band structure of Fe-doped 2D AlSb indicating contributions from Al-s, Al-p\textsubscript{x}, Al-p\textsubscript{y} and Al-p\textsubscript{z} orbitals. From this figure, contributions from Al-s orbital was negligible in comparison with Al-p orbital. And Al-p\textsubscript{z} orbital had the maximum contribution in the conduction band. Calculations of \autoref{pristine4201} were performed in Quantum ESPRESSO (QE). That's why the Fermi level was slightly shifted from valence band minima. There was also a small discrepancy in the band gap of Fe-doped 2D AlSb due to different algorithms of QE.

\clearpage
\begin{figure*}[h]
%\hspace{-2.70cm}
\centering
\includegraphics[width=\textwidth]{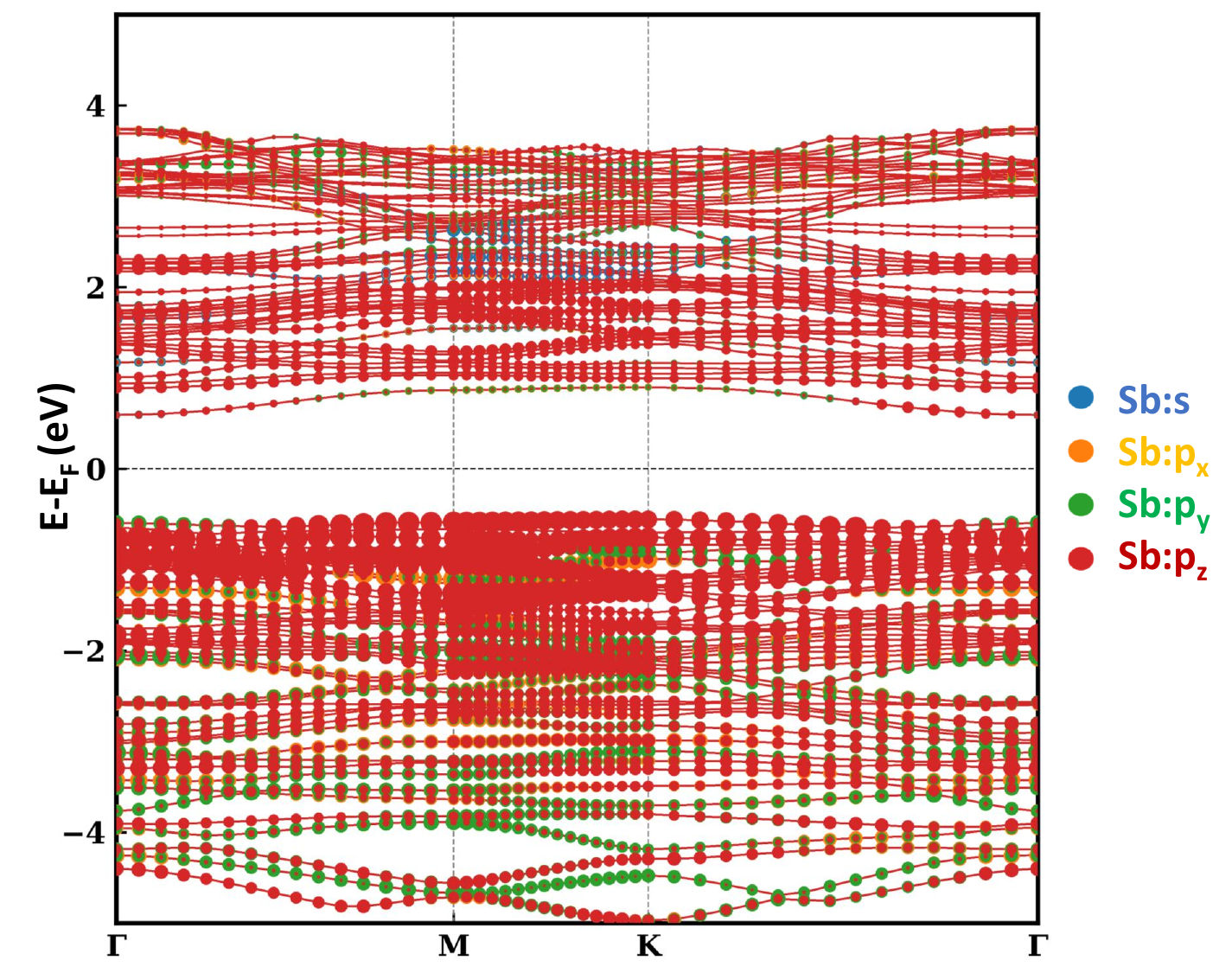}
  \caption{Orbital projected band structure of Fe-doped 2D AlSb, highlighting the contributions from Sb-s and Sb-p orbitals along the high-symmetry k-path. The circle size represents the orbital weight.}
  \label{pristine4202}
\end{figure*}

\autoref{pristine4202} is showing the orbital projected band structure of Fe-doped 2D AlSb indicating contributions from Sb-s, Sb-p\textsubscript{x}, Sb-p\textsubscript{y} and Sb-p\textsubscript{z} orbitals. From this figure, contributions from Sb-s orbital was negligible in comparison with Sb-p orbital. And Sb-p\textsubscript{z} orbital had the maximum contribution in the valence band. Whereas Al-p\textsubscript{z} contributed maximum in the conduction band previously shown in \autoref{pristine4201}.

\clearpage
\begin{figure*}[h]
%\hspace{-2.70cm}
\centering
\includegraphics[width=\textwidth]{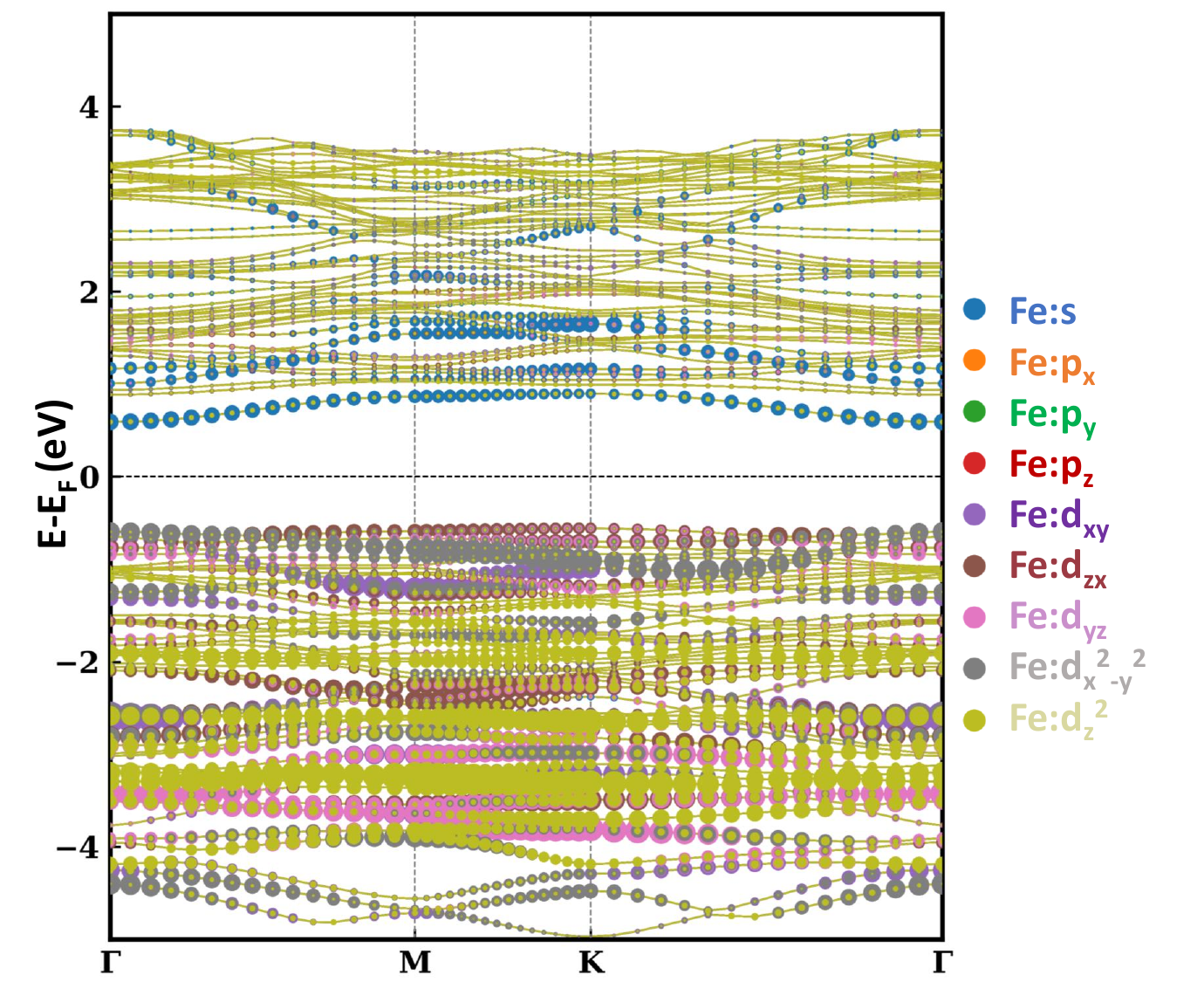}
  \caption{Orbital projected band structure of Fe-doped 2D AlSb, highlighting the contributions from Fe-s, Fe-p and Fe-d orbitals along the high-symmetry k-path. The circle size represents the orbital weight.}
  \label{pristine4203}
\end{figure*}
\autoref{pristine4203} is showing the orbital projected band structure of Fe-doped 2D AlSb indicating contributions from Fe-s, Fe-p\textsubscript{x}, Fe-p\textsubscript{y}, Fe-p\textsubscript{z}, Fe-d\textsubscript{xy}, Fe-d\textsubscript{zx}, Fe-d\textsubscript{yz}, Fe-d\textsubscript{x\textsuperscript{2}-y\textsuperscript{2}} and Fe-d\textsubscript{z\textsuperscript{2}} orbitals. From this figure, contributions from Fe-p orbitals were negligible in comparison with Fe-s and Fe-d orbitals. As Fe-s is the outermost orbital, that's why it had the highest contribution in the conduction band. Whereas Fe-d\textsubscript{zx}, Fe-d\textsubscript{x\textsuperscript{2}-y\textsuperscript{2}} and Fe-d\textsubscript{z\textsuperscript{2}} orbitals had the maximum contribution in the valence band.

\clearpage
\begin{figure*}[h]
%\hspace{-2.70cm}
\centering
\includegraphics[width=\textwidth]{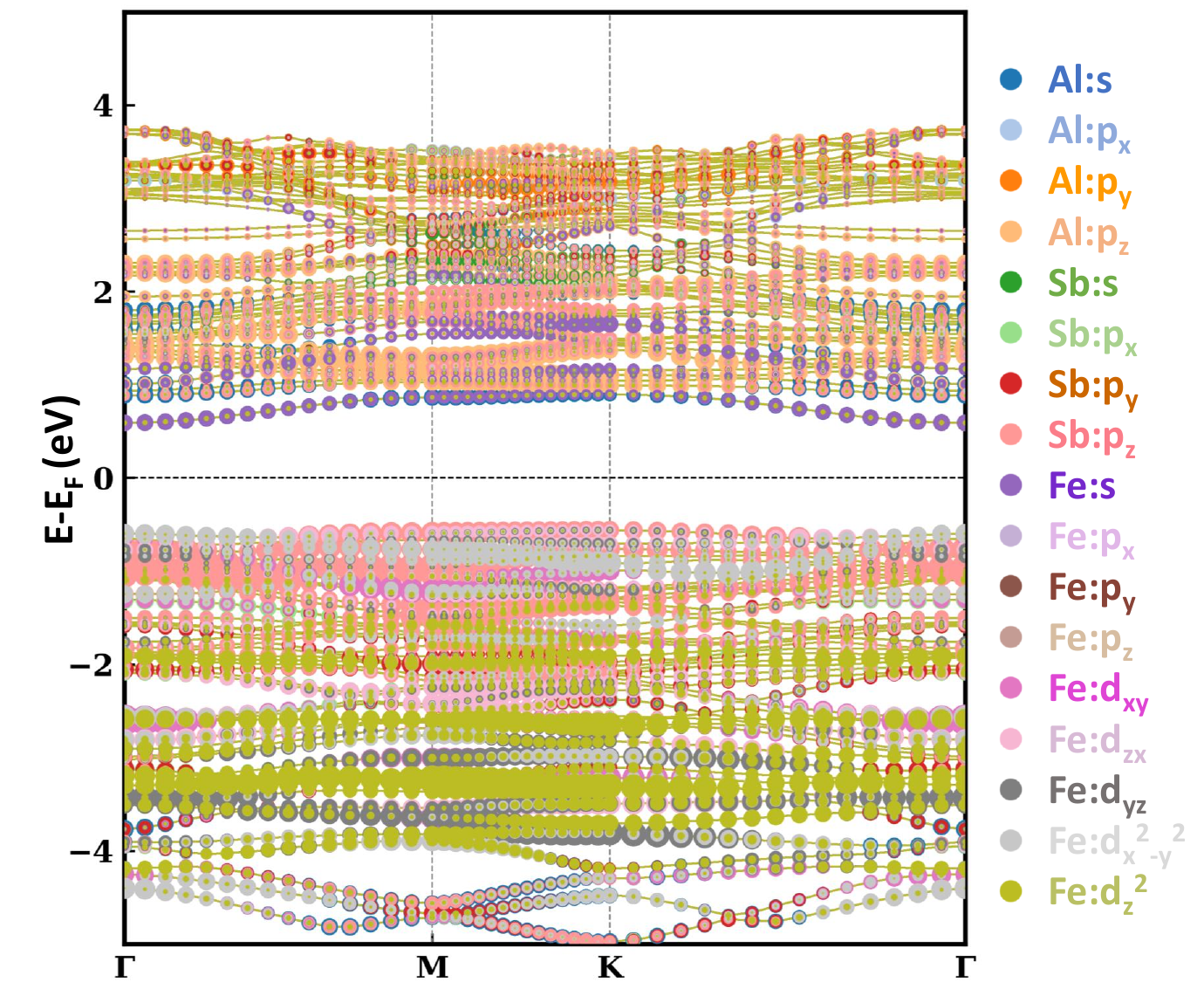}
  \caption{Orbital projected band structure of Fe-doped 2D AlSb, highlighting the contributions from the orbitals of Al, Sb and Fe simultaneously along the high-symmetry k-path. The circle size represents the orbital weight.}
  \label{pristine4204}
\end{figure*}
\autoref{pristine4204} is showing the orbital projected band structure of Fe-doped 2D AlSb indicating contributions from all the outermost orbitals of Al, Sb and Fe contemporaneously. From this figure, valence band was completely dominated by Fe-d\textsubscript{zx}, Fe-d\textsubscript{x\textsuperscript{2}-y\textsuperscript{2}} and Fe-d\textsubscript{z\textsuperscript{2}}. Whereas Fe-s and Sb-p\textsubscript{z} had the highest contributions in the conduction band. Also, weights of Fe-d orbitals were more extensive than that of other orbitals which indicated firmly that Fe-3d orbitals' interactions were the main reason for Fe doped 2D AlSb to be the most efficient CO\textsubscript{2}RR catalyst studied in this paper.

\clearpage
\section{Reaction pathways of Co- and Ni-doped 2D AlSb}
\begin{figure*}[h]
%\hspace{-2.70cm}
\centering
\includegraphics[width=\textwidth]{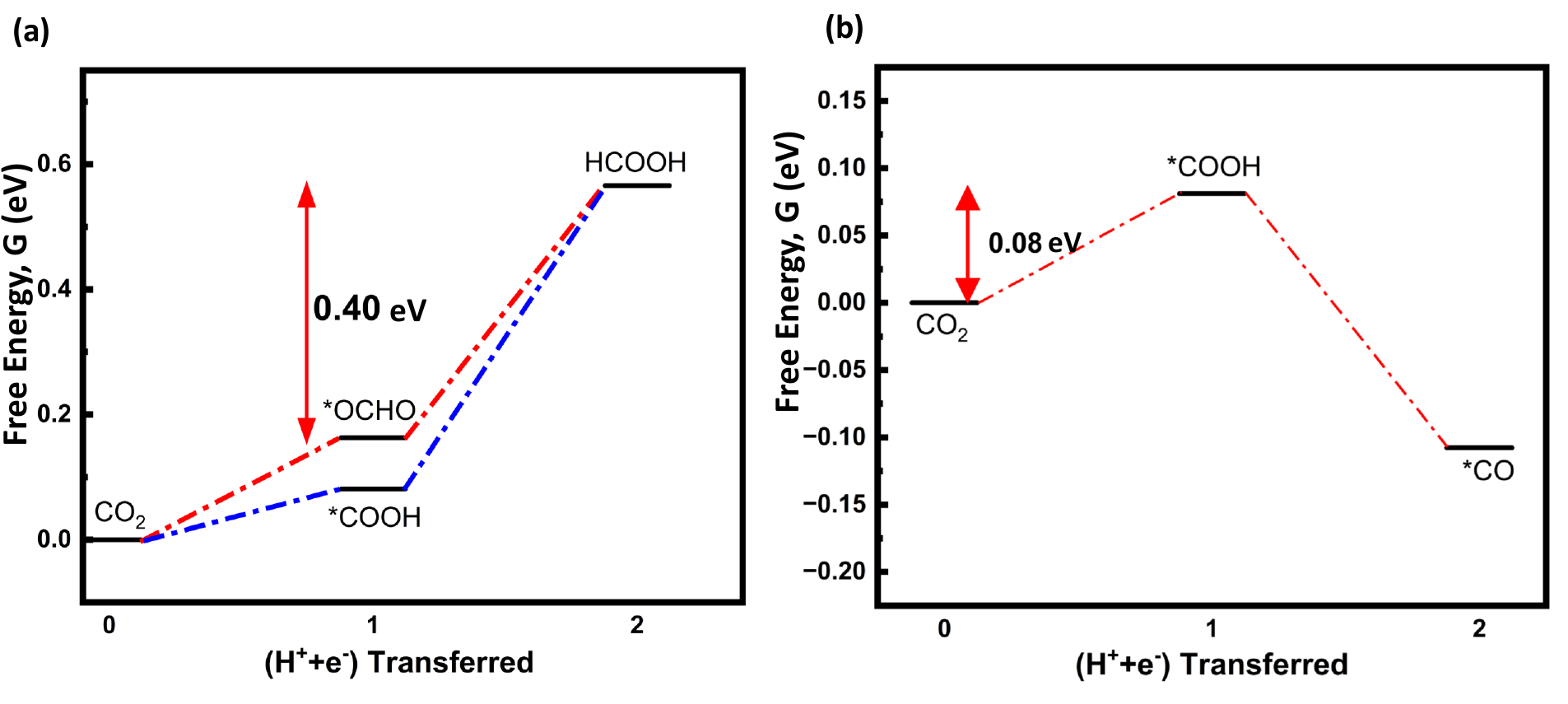}
\caption{Reaction pathway for CO\textsubscript{2} reduction on Co-doped 2D AlSb to (a)HCOOH, (b)CO.}
  \label{pristine622}
\end{figure*}

\begin{figure*}[h]
%\hspace{-2.70cm}
\centering
\includegraphics[width=\textwidth]{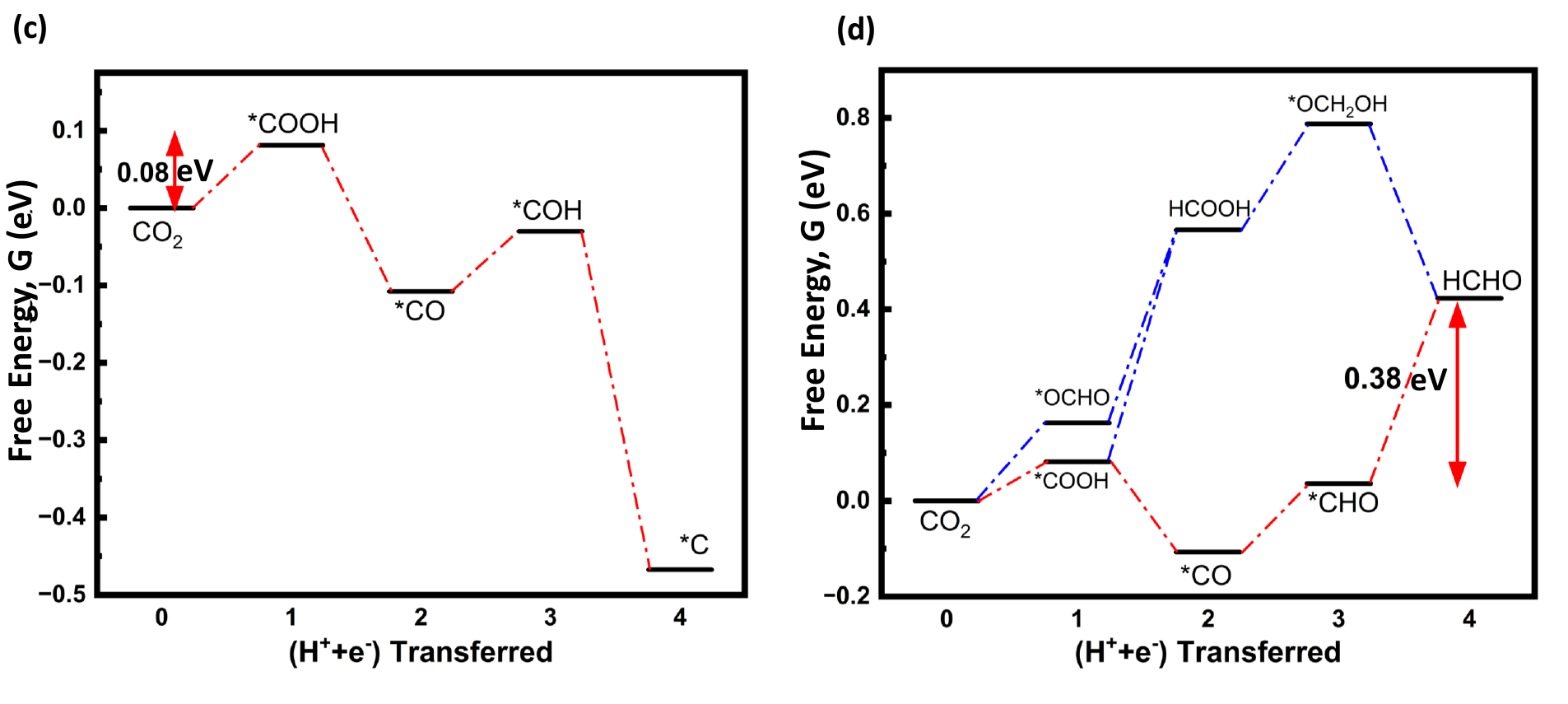}
\caption{Reaction pathway for CO\textsubscript{2} reduction on Co-doped 2D AlSb to (c)*C, (d)HCHO.}
  \label{pristine722}
\end{figure*}
\autoref{pristine622} is showing reaction pathways to HCOOH and *CO for Co-doped 2D AlSb. For pathway to HCOOH, *OCHO to HCOOH was the potential determining step (PDS) and overpotential from CO\textsubscript{2} to HCOOH conversion for Co-doped 2D AlSb was 0.40 eV. Similarly, for pathway to *CO, CO\textsubscript{2} to *COOH was the PDS and overpotential from CO\textsubscript{2} to *CO conversion for Co-doped 2D AlSb was 0.08 eV. And \autoref{pristine722} is showing reaction pathways to *C and HCHO for Co-doped 2D AlSb. For pathway to *C, CO\textsubscript{2} to *COOH was the PDS and overpotential from CO\textsubscript{2} to *C conversion for Co-doped 2D AlSb was 0.08 eV. Similarly, for pathway to HCHO, *CHO to HCHO was the PDS and overpotential from CO\textsubscript{2} to HCHO conversion for Co-doped 2D AlSb was 0.38 eV.

\clearpage
\begin{figure*}[h]
%\hspace{-2.70cm}
\centering
\includegraphics[width=\textwidth]{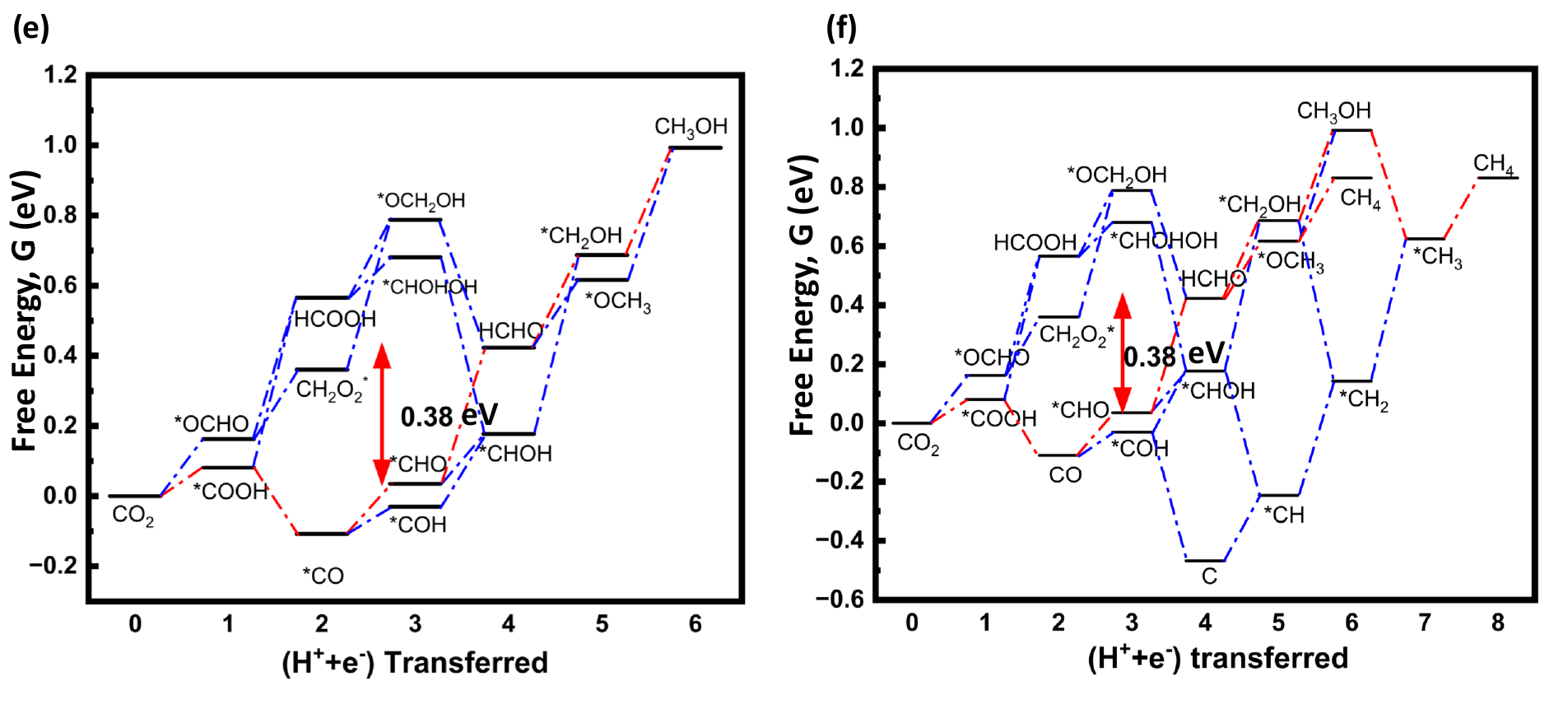}
\caption{Reaction pathway for CO\textsubscript{2} reduction on Co-doped 2D AlSb to (e)CH\textsubscript{3}OH, (f)CH\textsubscript{4}.}
  \label{pristine832}
\end{figure*}

\begin{figure*}[h]
%\hspace{-2.70cm}
\centering
\includegraphics[width=\textwidth]{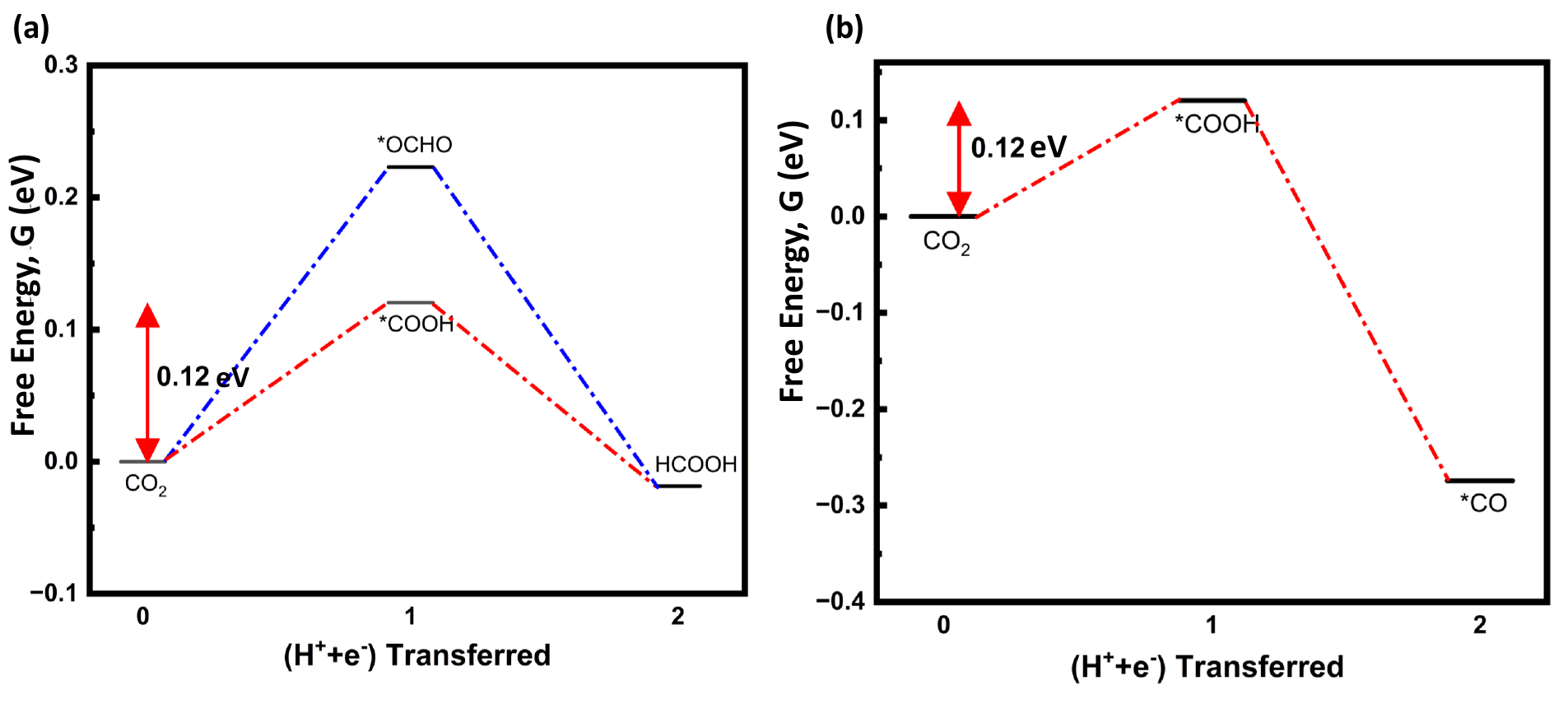}
\caption{Reaction pathway for CO\textsubscript{2} reduction on Ni-doped 2D AlSb to (a)HCOOH, (b)CO.}
  \label{pristine932}
\end{figure*}
\autoref{pristine832} is showing reaction pathways to CH\textsubscript{3}OH and CH\textsubscript{4} for Co-doped 2D AlSb. For pathway to CH\textsubscript{3}OH, *CHO to HCHO was the PDS and overpotential from CO\textsubscript{2} to CH\textsubscript{3}OH conversion for Co-doped 2D AlSb was 0.38 eV. Similarly, for pathway to CH\textsubscript{4}, *CHO to HCHO was the PDS and overpotential from CO\textsubscript{2} to CH\textsubscript{4} conversion for Co-doped 2D AlSb was 0.38 eV. And \autoref{pristine932} is showing reaction pathways to HCOOH and *CO for Ni-doped 2D AlSb. For pathway to HCOOH, CO\textsubscript{2} to *COOH was the PDS and overpotential from CO\textsubscript{2} to HCOOH conversion for Ni-doped 2D AlSb was 0.12 eV. Similarly, for pathway to *CO, CO\textsubscript{2} to *COOH was the PDS and overpotential from CO\textsubscript{2} to *CO conversion for Ni-doped 2D AlSb was 0.12 eV.

\clearpage

\begin{figure*}[h]
%\hspace{-2.70cm}
\centering
\includegraphics[width=\textwidth]{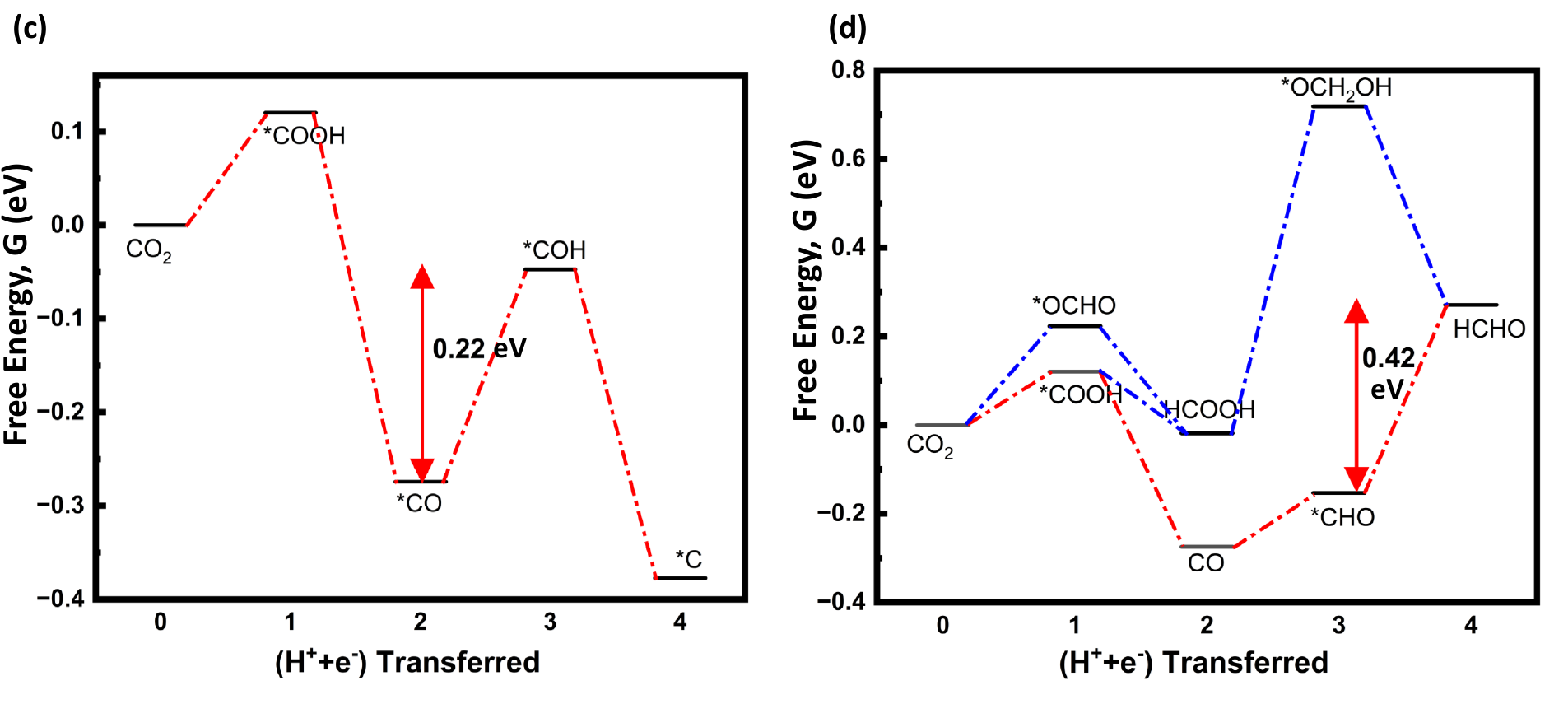}
\caption{Reaction pathway for CO\textsubscript{2} reduction on Ni-doped 2D AlSb to (c)*C, (d)HCHO.}
  \label{pristine10}
\end{figure*}

\begin{figure*}[h]
%\hspace{-2.70cm}
\centering
\includegraphics[width=\textwidth]{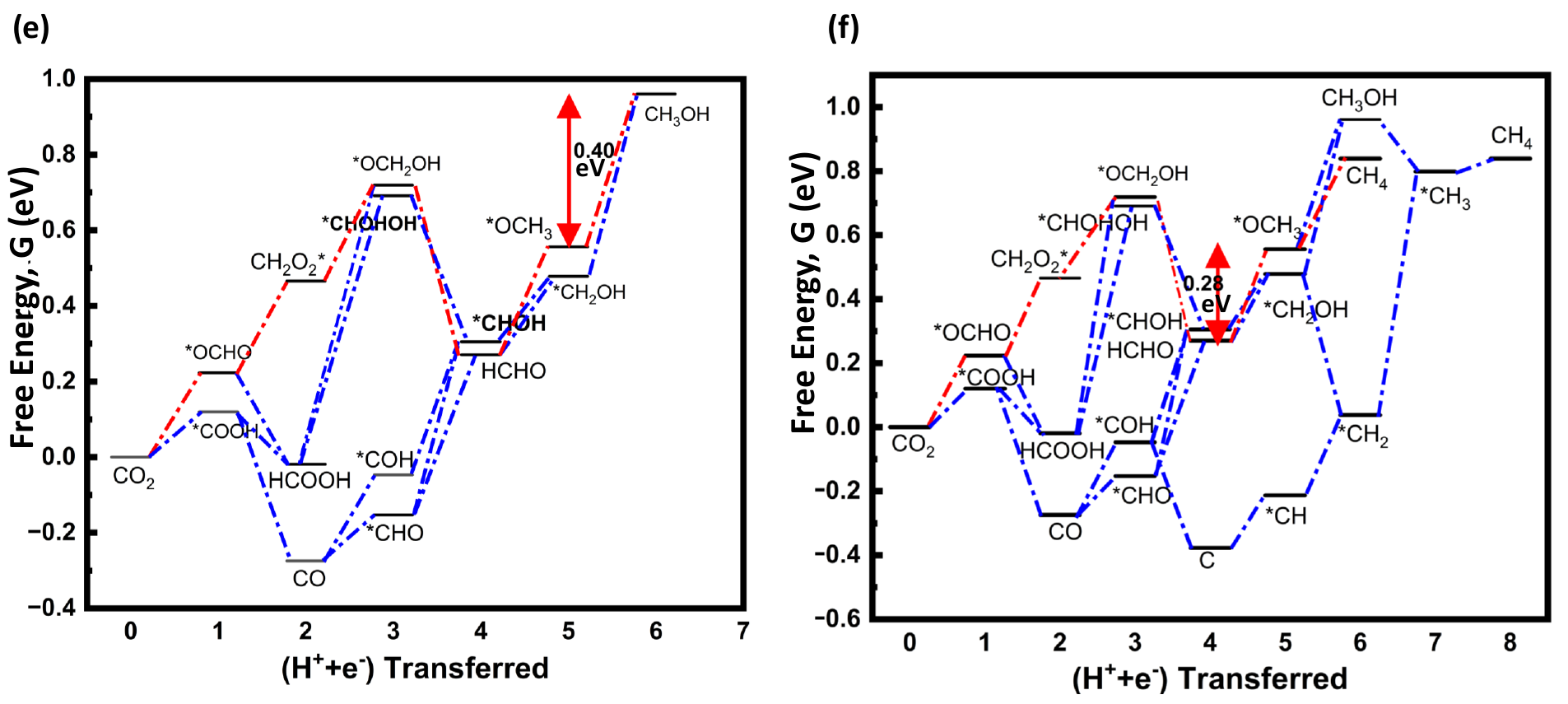}
\caption{Reaction pathway for CO\textsubscript{2} reduction on Ni-doped 2D AlSb to (e)CH\textsubscript{3}OH, (f)CH\textsubscript{4}.}
  \label{pristine11}
\end{figure*}

\autoref{pristine10} is showing reaction pathways to *C and HCHO for Ni-doped 2D AlSb. For pathway to *C, *CO to *COH was the PDS and overpotential from CO\textsubscript{2} to *C conversion for Ni-doped 2D AlSb was 0.22 eV. Similarly, for pathway to HCHO, *CHO to HCHO was the PDS and overpotential from CO\textsubscript{2} to HCHO conversion for Ni-doped 2D AlSb was 0.42 eV. And \autoref{pristine11} is showing reaction pathways to CH\textsubscript{3}OH and CH\textsubscript{4} for Ni-doped 2D AlSb. For pathway to CH\textsubscript{3}OH, *OCH\textsubscript{3} to CH\textsubscript{3}OH was the PDS and overpotential from CO\textsubscript{2} to CH\textsubscript{3}OH conversion for Ni-doped 2D AlSb was 0.40 eV. Similarly, for pathway to CH\textsubscript{4}, HCHO to *OCH\textsubscript{3} was the PDS and overpotential from CO\textsubscript{2} to CH\textsubscript{4} conversion for Ni-doped 2D AlSb was 0.28 eV.

\clearpage
\section{Gibbs free energy calculation under applied potential}
\begin{figure*}[h]
%\hspace{-2.70cm}
\centering
\includegraphics[height=0.50\linewidth]{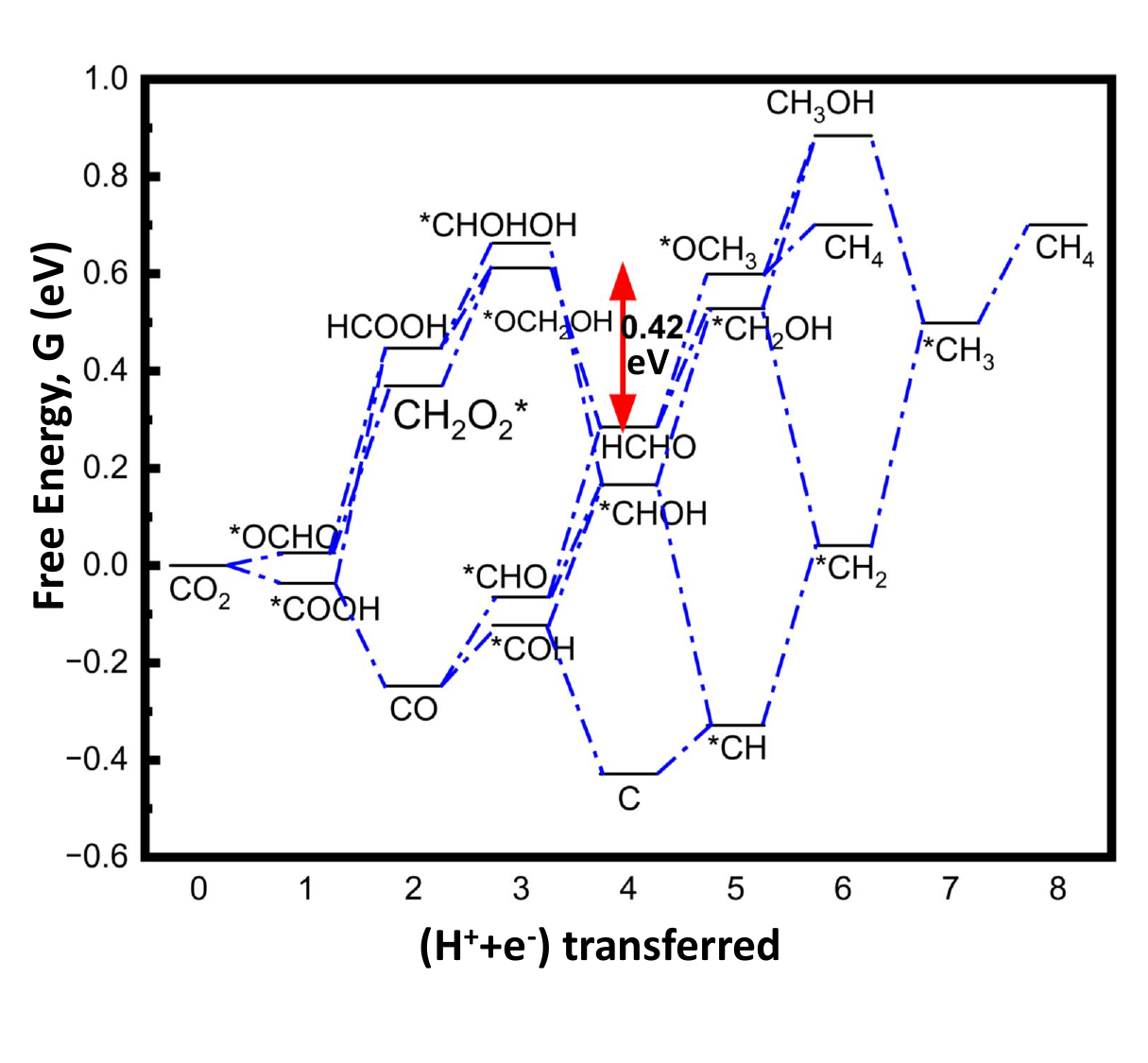}
  \caption{
  Free energy pathway to 8e product CH$_4$ on pristine 2D AlSb.
  }
  \label{pristine1212}
\end{figure*}

\begin{figure*}[h]
%\hspace{-2.70cm}
\centering
\includegraphics[height=0.48\linewidth]{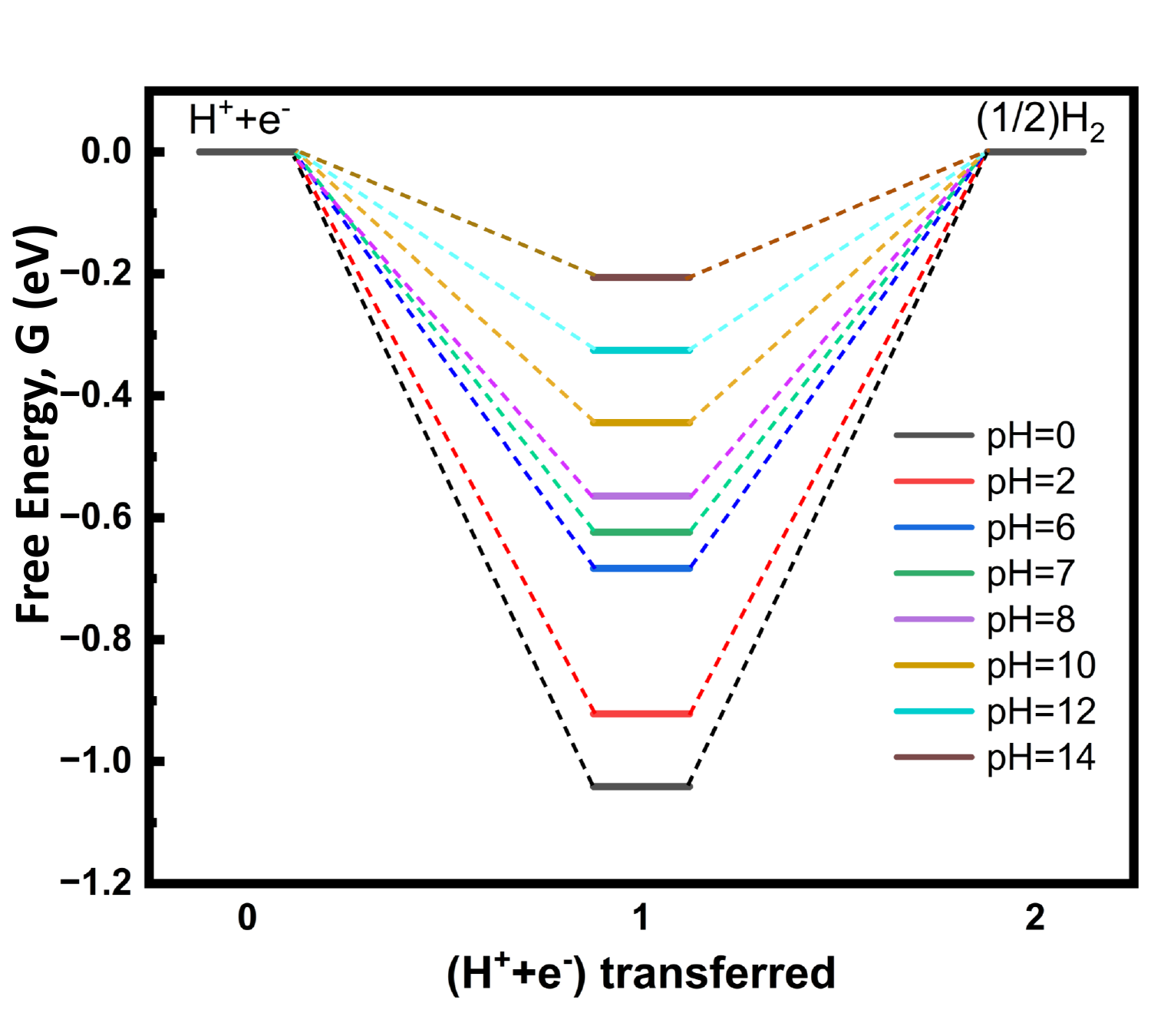}
  \caption{Free energy diagram of HER on pristine 2D AlSb with pH variation.}
  \label{pristine1313}
\end{figure*}
From \autoref{pristine1313} of hydrogen evolution reaction (HER) on pristine 2D AlSb, we have \( \Delta G_{\text{HER}} = -1.0412 \) eV, indicating that pristine 2D AlSb exhibits less catalytic activity for CO$_2$RR under acidic conditions (\( pH = 0 \)). From \autoref{pristine1212} of CO$_2$RR for conversion from CO$_2$ to CH$_4$ on pristine 2D AlSb,  
\[
\Delta G_{\text{Max}} = 0.421200537 \text{ eV}
\]

\begin{figure*}[h]
%\hspace{-2.70cm}
\centering
\includegraphics[height=0.48\linewidth]{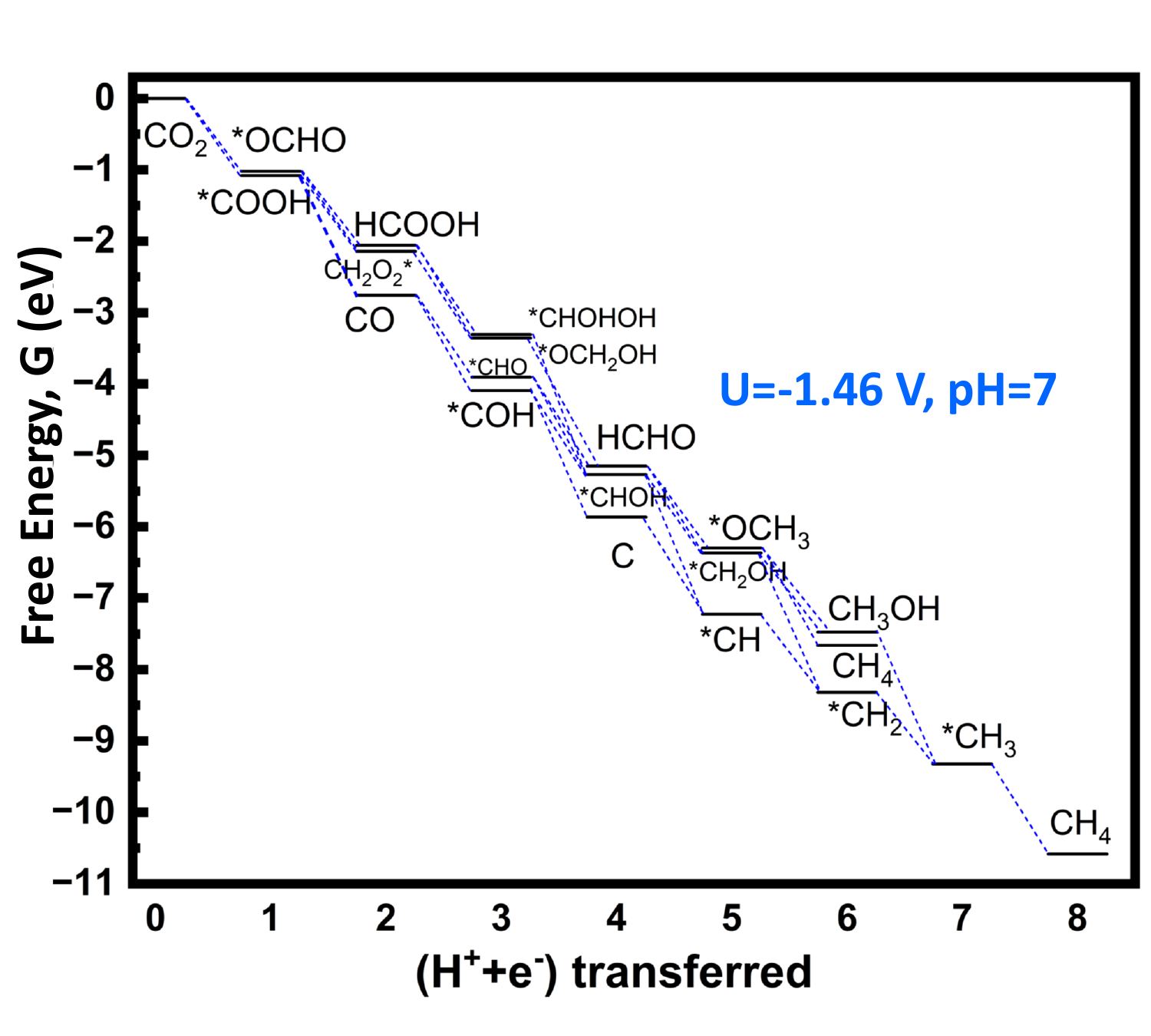}
  \caption{Free energy pathway to 8e product CH$_4$ on pristine 2D AlSb with applied potential and pH.}
  \label{pristine1414}
\end{figure*}

At pH=7 (neutral medium) and applied potential U (volt), the condition for CO\textsubscript{2}RR be more likely than for HER, 
\[
\Delta G_{\text{HER}} > \Delta G_{\text{Max,old}} - neU
\]

\[
-1.0412 > 0.421200537 - neU, \quad n = 1
\]

\[
|U| > 1.463 \text{ V}, \quad n = 1
\]
The modified pathway after the applied potential and pH is shown in \autoref{pristine1414}. Similar calculations were performed for Fe-, Co- and Ni-doped 2D AlSb.

\end{document}